\documentclass[aps,prb,groupedaddress,twocolumn]{revtex4-1}

\usepackage{graphicx}
\usepackage{amsmath}
\usepackage{braket}
\usepackage{appendix}

\begin{document}


\title{An alternative functional renormalization group approach to the single impurity Anderson model}


\author{Michael Kinza$^{1}$}
\email[]{kinza@physik.rwth-aachen.de}
\author{Jutta Ortloff$^{2}$}
\author{Johannes Bauer$^{3,4}$}
\author{Carsten Honerkamp$^{1}$}
\affiliation{
$^1$ Institute for Solid State Theory, RWTH Aachen University,
D-52056 Aachen and JARA - Fundamentals of Future Information
Technology\\
$^2$ Theoretical Physics, University of W\"{u}rzburg, D-97074 W\"{u}rzburg, Germany\\
$^3$ Max-Planck Insititute for Solid State Research, Heisenbergstr.1, 70569 Stuttgart, Germany\\
${}^4$ Department of Physics, Harvard University, Cambridge,
  Massachusetts 02138, USA}

\date{\today}

\begin{abstract}
We present an alternative functional renormalization group (fRG) approach to
the single-impurity Anderson model at finite temperatures. Starting
with the exact self-energy and interaction vertex of a small system
('core') containing a correlated site, we switch on the hybridization with a
non-interacting bath in the fRG-flow and calculate spectra of the correlated
site. Different truncations of the RG-flow-equations and choices of the core are
compared and discussed. Furthermore we calculate the linear conductance and the magnetic susceptibility as functions
of temperature and interaction strength. The signatures of Kondo physics arising in the flow are compared with
numerical renormalization group results. 
\end{abstract}

\pacs{}

\maketitle


\section{Introduction}

The single-impurity Anderson model (SIAM) is a minimal model to describe the
interplay of charge and spin-fluctuations of an interacting impurity in a
metallic environment, including the Kondo effect. Through decades of
theoretical research since its first 
proposal\cite{And61} in the 1960s  it has been thoroughly investigated. There
are exact solutions from the Bethe-Ansatz-technique, \cite{Hew93,Andr83} and
an accurate method to describe static and dynamic 
properties is Wilson's numerical renormalization group (NRG). \cite{Wil75,Bul08}

In the last twenty years renewed interest arose in analyzing the SIAM. One
reason is the fabrication of nanoscale devices, in which quantum dots are
coupled to metallic leads. In certain cases they can be described by the SIAM
and the Kondo effect was observed. \cite{Gol97} Another reason is the development of
the dynamical mean field theory (DMFT). \cite{Met89,Geo96} In the latter a lattice model
is mapped to an impurity model coupled to a dynamical Weiss-field bath that has to be
determined selfconsistently. 
Therefore the theoretical challenge remains to develop versatile and
numerically inexpensive methods that can describe a large class of impurity
models appearing in this context. The single channel SIAM can serve as  
a benchmark to test those methods. One approach to tackle the impurity problems that
has been developed is the functional renormalization group (fRG).
\cite{Met12} Even though the fundamental equation of this framework is exact
most methods based on the fRG are perturbative. Hence, so far it has been difficult to accurately
resolve the non-perturbative Kondo physics. However, the transparency and
flexibility of the fRG can lead to useful 
applications in more complex contexts, where, for instance, the NRG is difficult to apply.    
fRG-approaches to the Anderson impurity model come already in some variety,
e.g. there are variants based on a frequency cutoff, \cite{Kar06,Hed04,Kar08}
on Hubbard-Stratonovich-fields representing spin-fluctuations
\cite{Bar09,Isi10,Stre12} and on a flowing level broadening. \cite{Jak10} Also,
non-equilibrium situations are subject to current research.
\cite{Gez07,Jak07,Kar10b} 

Here we introduce and test another fRG approach to the SIAM. As opposed to the
previous approaches, our method starts with the exact solution of a small
system of a few sites, which is termed the 'core'. The fRG flow then couples
the core adiabatically to a bath of non-interacting fermions, in a 'hybridization flow'.  
The main motivation for this approach is the following. The usual hierarchy of
fermionic fRG equations for the fermionic vertex functions has to be truncated
by neglecting the higher-order vertex functions, typically after the
four-point vertex. In the usual context without bare higher-order interactions
and in standard perturbation theory, these higher-order vertices would come in
higher orders in the bare interactions. Hence, the expectation is that the
truncation can only be good at weaker interactions. For a normal many-fermion
system with a full Fermi surface, in the beginning of the fRG flow, the
higher-order terms are suppressed by these higher orders of the bare
interactions, while at low scales, near the Fermi surface additional phase
space arguments may limit their impact. For strong initial interaction no
argument can be given that the impact of these neglected vertex functions is
negligible. Another expectation is however that these higher-order terms are
mainly determined by local physics and by degrees of freedom over a larger
energy range in terms of the free Hamiltonian. Therefore one may hope to
arrive at a satisfactory description also for stronger interaction by incorporating the higher-order
vertices of only a small system and by neglecting their change when the
low-energy physics is altered during an fRG flow. Hence, in the present
approach, we use the exact four-point vertex and self-energy of a small system
as starting point for the hybridization flow. These quantities have built in
the effect of all orders in the interaction at least for this small system. Now,
performing the truncated fRG flow, the hybridization-induced change of the back-effect of the
higher-order interactions on the four-point and ultimately on the self-energy
will be missing, but this may still be better than ignoring the higher-order
physics completely. 

Note that this strategy which we are testing here for an impurity problem,
could also be extended to a lattice problem. One can imagine
using the small-cluster self-energy and 
four-point-vertices as initial condition for a flow in the band width or
hopping amplitude of a lattice dispersion. Similar strategies have already
been pursued for bosonic problems. \cite{Ran11,Ran11b} In this context, the
present study can be seen as first step in the exploration of such a procedure
for fermions, with the benefit that in impurity models quantitative 
benchmarking is possible.  

The application of RG flow equations usually requires a controlled starting point in
the parameter space of the theory where the vertices are well known. Then one
can follow the flow toward a nontrivial physical point of the theory. In our case the
flow takes place in the effective theory of the first bath site next to the
impurity or correlated core system. Initially, the bath is decoupled, and
hence the bath site is non-interacting, providing a well-defined starting
point with finite density of states at low energies. Then, in the RG flow, the
coupling to the correlated core is switched on and increased to the desired
value. Thereby the bath theory becomes interacting and the spectrum of the
bath sites is modified. Employing exact relations of the first bath-site
self-energy to the self energy of the correlated core, we can then deduce the
spectrum of the correlated core as well and study the signatures of Kondo
physics. 

This paper is organized as follows. In Section \ref{sec:siam}, we describe the
single-impurity Anderson model and its Green's functions on the correlated
site and on neighboring sites. In Sec. \ref{sec:effbath}, we derive the effective
bath theory on the first bath site and give relations between the bath
self-energies and the self-energy of the correlated site. In
Sec. \ref{sec:FRGeqs}, we describe the fRG scheme in the effective bath
theory. Sec. \ref{sec:results} is the devoted to numerical results. We
conclude with a discussion in Sec. \ref{sec:conclusions}. 
 
\section{The Single Impurity Anderson Model}
\label{sec:siam}

\subsection{Hamiltonian}
The Hamiltonian of the single channel SIAM consists of three parts
\begin{equation}\label{eqhamiltonsiam}
\hat{H} = \hat{H}_{dot} + \hat{H}_{bath} + \hat{H}_{bath-dot} \, . 
\end{equation}
$\hat{H}_{dot}$ describes the interacting electron level and is given by
\begin{equation}
\hat{H}_{dot} = \sum_{\sigma}\left(\epsilon_{d,\sigma}-\mu\right)d^{\dag}_{\sigma}d_{\sigma} + U \ d^{\dag}_{\uparrow}d_{\uparrow}d^{\dag}_{\downarrow}d_{\downarrow} \, . 
\end{equation} 
The operators $d^{\dag}_{\sigma}$ and $d_{\sigma}$ create and annihilate electrons on the dot-level with spin-component $\sigma = \pm 1$.
The onsite-energy is given by
\begin{equation}
\epsilon_{d,\sigma}-\mu=-\frac{U}{2}+V_{g}+B\sigma \, .
\end{equation}
including a magnetic field term $B=g\mu_{\rm B}H$ with Bohr magneton $\mu_{\rm
  B}$ and a gate-voltage energy $V_{g}$. The term 
$-\frac{U}{2}$ is choosen such that $V_{g}=0$ corresponds to the
particle-hole-symmetric point.

Our bath consists of two semi-infinite tight-binding-chains with hopping-parameter $t$
\begin{eqnarray}
\nonumber
\hat{H}_{bath} &=& -t \sum_{s=L,R}\sum_{\sigma}\sum_{j=1}^{\infty}\left( b^{\dag}_{j,\sigma,s}b_{j+1,\sigma,s}+H.c.\right)\\
&&-\mu \sum_{s=L,R}\sum_{\sigma}\sum_{j=1}^{\infty} b^{\dag}_{j,\sigma,s}b_{j,\sigma,s} \,  .
\end{eqnarray}
The operators $b^{\dag}_{j,\sigma,s}$ and $b_{j,\sigma,s}$ create and
annihilate electrons on the site $j$ of the left ($s=L$) or the right ($s=R$) 
bath with spin-component $\sigma$. Different choices of the bath could be easily
incorporated into our formalism. 
The coupling between the bath and the interacting dot-site is given by
\begin{equation}
\hat{H}_{bath-dot} = -\overline{v}\sum_{s=L,R}\sum_{\sigma}\left(b^{\dag}_{1,\sigma,s}d_{\sigma}+H.c.\right) \, . 
\end{equation}
We can do a unitary transformation
\begin{equation}
\left(\begin{array}{c} b_{j,\sigma,even} \\ b_{j,\sigma,odd} \end{array}\right) = \frac{1}{\sqrt{2}} \left(\begin{array}{cc} 1 & 1 \\ 1 & -1 \end{array}\right)
\left(\begin{array}{c} b_{j,\sigma,L} \\ b_{j,\sigma,R} \end{array}\right)
\end{equation}
such that only the even combination coupled to the dot site,
since the left and right part of the chain possess the same chemical potential,
\begin{equation}
\hat{H}_{bath-dot}=-v\sum_{\sigma}\left(b^{\dag}_{1,\sigma,even}d_{\sigma}+H.c.\right)
\, ,  
\end{equation}
where $v=\sqrt{2}\overline{v}$. $\hat{H}_{bath}$ remains formally unchanged.
In the following we ignore the decoupled odd bath and skip the index 'even' on
the remaining even bath. Furthermore we set $\mu=0$.

\subsection{Green's function of the SIAM}

The free Green's function on the dot-site is given by
\begin{equation}
\mathcal{G}^{0}_{\sigma}\left(i\omega_{n},d,d\right)=\left[i\omega_{n}-\epsilon_{d,\sigma}-\Delta\left(i\omega_{n}\right)\right]^{-1} \,
\end{equation}
where $\Delta\left(i\omega_{n}\right)$ is the hybridization function, which is given by
\begin{equation}
\Delta\left(i\omega_{n}\right)=v^2 g_{b}\left(i\omega_{n},b_{1},b_{1}\right) \,  
\end{equation} 
with $g_{b}\left(i\omega_{n}\right)=g_{b}\left(i\omega_{n},b_{1},b_{1}\right)$
\begin{equation}
\label{eqbadmatsubaragreenfunktion}
g_{b}\left(i\omega_{n}\right)=\frac{1}{2t^2}\left(i\omega_{n}-i
  \text{sgn}(\omega_{n})\sqrt{4t^2-(i\omega_{n})^2}\right) .
\end{equation}
Details for the derivation are given in appendix \ref{sec:appendixgreenfunction}.

The retarded Green's function on the first bath site is given by
\begin{eqnarray}
g_{b}\left(\omega+i 0^{+}\right)&=&\frac{1}{2t^2}\Big(
- i\sqrt{4t^2-\omega^2}\Theta\left(2t-|\omega|\right) \nonumber
\\
&+& \omega-
\sqrt{\omega^2-4t^2}\Theta\left(|\omega|-2t\right){\rm sgn}(\omega)\Big). \nonumber
\end{eqnarray}
The density of states on the first bath site is then semi-elliptic
\begin{eqnarray}
\nonumber
\rho_b\left(\omega\right)&=&-\frac{1}{\pi}\text{Im}\left[g_{b}\left(\omega+i
    0^{+}\right)\right] \nonumber\\ 
&=&\frac{1}{2\pi t^2}\sqrt{4 t^2-\omega^2}\Theta\left(2t-|\omega|\right) ,
\end{eqnarray}
with band width $W=4t$. In most studies of the SIAM in the literature one considers a constant DOS and
the wide-band limit, i.e., $W$ is much larger than all other scales for the problem. Then
the physics for the symmetric model SIAM mostly depends on the ratio of
interaction scale $U$ and the hybridization scale $\Delta$. Here we keep the
$\omega$-dependence of the hybridization function. We define the quantity $\Delta_0=\pi v^2
\rho_b\left(0\right)=\frac{v^2}{t}$. We choose for
simplicity $v=t$, so that $\Delta_0 = t$. This means that we do not have two
independent parameters for bandwidth and hybridization as one usually does for
studies of the SIAM and the finite bandwidth actually enters the
problem. Therefore, our results differ quantitatively from the the wide-band
limit, which is common in the literature. In some sense it is more similar to
the first iteration of a DMFT calculation with a semi-elliptic DOS. 
We take $\Delta_0=t=1$ as reference energy scale in the following.
We would like to emphasize again at this point that our formalism can also
deal with more general bath functions, as long as they can be mapped onto a
linear chain with certain onsite and hopping parameters.

Because the bath is noninteracting, the self-energy is local on the dot
site. By the Dyson equation the full Green's function 
is related to $Q_{\sigma}\left(i\omega_{n}\right)$ and the full Green's
function on the dot reads
\begin{equation}
\label{eqfullgreensfunktiondotsite}
\mathcal{G}_{\sigma}\left(i\omega_{n},d,d\right)=
\left[i\omega_{n}-\epsilon_{d,\sigma}-\Sigma_{d,\sigma}\left(i\omega_{n}\right)
-\Delta\left(i\omega_{n}\right)\right]^{-1}. 
\end{equation}

\section{Effective theory for the bath}
\label{sec:effbath}
\subsection{Integrating out the 'core'}

\begin{figure}[htbp]
\vspace*{0.4cm}
    \centering
    \includegraphics[width=0.40\textwidth]{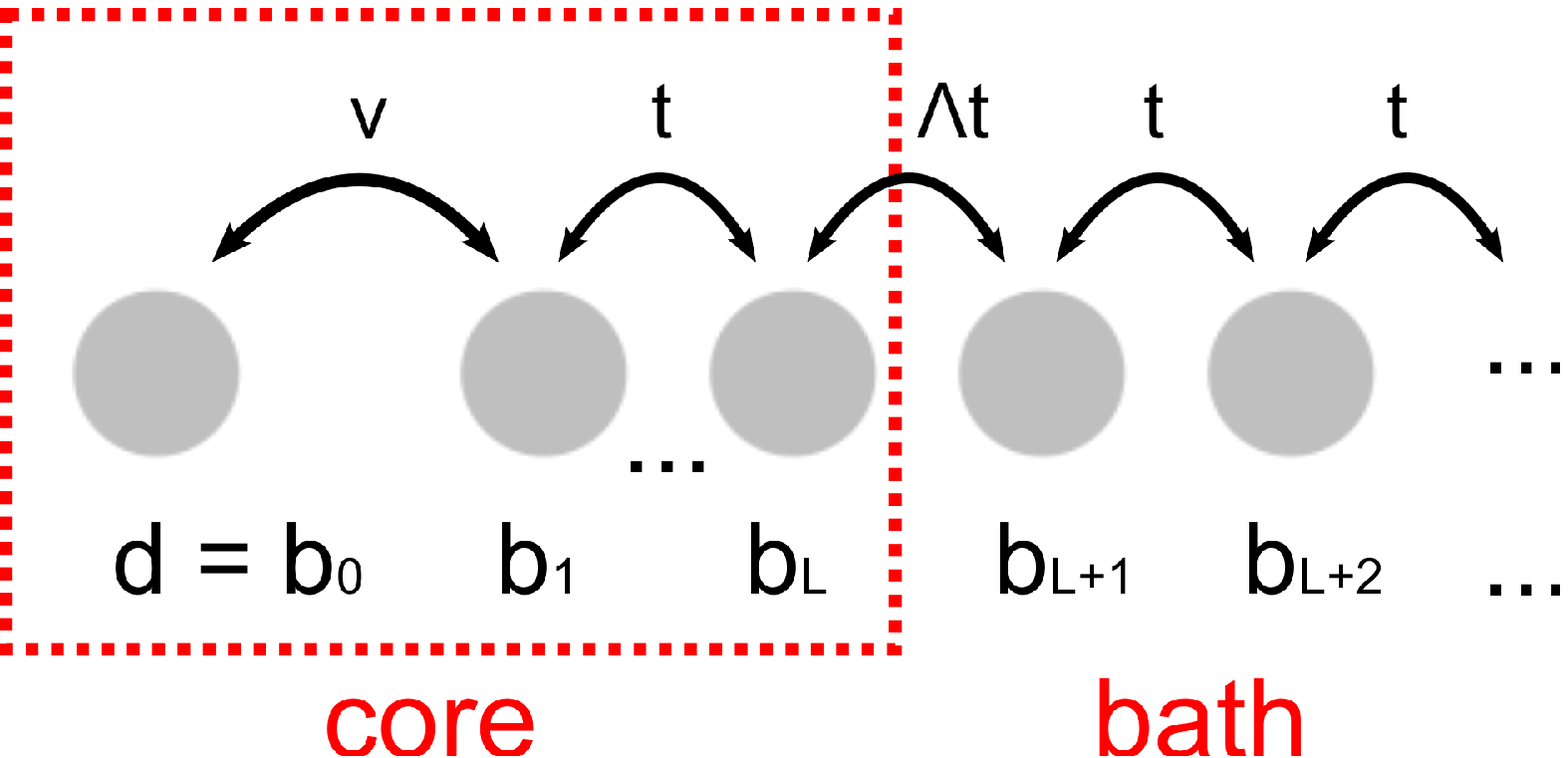}
    \caption[andersonmodel]{The semi-infinite tight-binding-chain is separated
      into two parts: The 'core' includes the correlated site and L
      bath-sites. The 'bath' consists of the remaining bath-sites. The two
      parts are coupled by a hopping-term which is proportional to the
      parameter $\Lambda$.} 
    \label{pictandersonmodel}
\end{figure}

We now separate the system into two parts as illustrated in Fig.
\ref{pictandersonmodel}. One part (called 'core' in the following) contains
the correlated site and the first $L$ bath-sites of the noninteracting 
tight-binding chain ($L=0,1,2,3$). The other part (called 'bath') contains all
bath-sites of the tight-binding-chain with index $i \geq L+1$. 
In the following we integrate out the 'core' in a functional integral
representation of our model leading to an effective theory for the bath.  
\footnote{The idea of integrating out the correlated site to reduce the
  Anderson model to an effective bath-theory is for the case $L=0$ worked out
  in Ref. \onlinecite{Joe10}. In this work analytic results that are
  perturbative in the effective bath-interaction as well as fRG-results are
  presented. These served as a benchmark to our numerical results.} 

Our model is described by the grandcanonical partition function
\begin{equation}
\mathcal{Z}=\int\mathcal{D}\left[\bar{b},b\right]\ \exp\left[-S\left(\bar{b},b\right)\right]
\end{equation}
with the action
\begin{eqnarray}\label{eqmicroscopicaction}
\nonumber
S\left(\bar{b},b\right)&=&S_{core}\left(\bar{b}_0,b_0,\bar{b}_{1},b_{1},...,\bar{b}_{L},b_{L}\right)\\
\nonumber
&+& S_{bath}\left(\bar{b}_{L+1},b_{L+1},\bar{b}_{L+2},b_{L+2},...\right)\\
&+& S_{coupling}^{\Lambda}\left(\bar{b}_{L},b_{L},\bar{b}_{L+1},b_{L+1}\right) \, . 
\end{eqnarray}
To make the notation more compact, we denote the dot-site $d$ as the 0th
bath-site $b_0$. We define the scalar product
\begin{equation}
\left(\psi,\phi\right)=\int_{0}^{\beta}d\tau \sum_{\sigma}\psi_{\sigma}(\tau)\phi_{\sigma}(\tau) .
\end{equation}
$S_{core}$ and $S_{bath}$ are given by
\begin{widetext}
\begin{eqnarray}
\label{eqeffectiveactioncore}
S_{core}\left(\bar{b}_{0},b_{0},\bar{b}_{1},b_{1},...,\bar{b}_{L},b_{L}\right)
&=&\left(\bar{b}_{0},(\partial_{\tau}+\epsilon_{d,\sigma})b_{0}\right)
 +U\  \int_{0}^{\beta}d\tau\ n_{\uparrow}\left(\tau\right)n_{\downarrow}\left(\tau\right)\\
\nonumber
&& -v\ \left[\left(\bar{b}_{0},b_{1}\right)+H.c.\right]
+ \sum_{j=1}^{L}\left(\bar{b}_{j},\partial_{\tau}b_{j}\right)
-t\ \sum_{j=1}^{L-1}\left[\left(\bar{b}_{j},b_{j+1}\right)+H.c.\right] \,  ,\\[1mm]
S_{bath}\left(\bar{b}_{L+1},b_{L+1},\bar{b}_{L+2},b_{L+2},...\right)
&=&  \sum_{j=L+1}^{\infty} \left(\bar{b}_{j},\partial_{\tau}b_{j}\right)
-t\ \sum_{j=L+1}^{\infty}\left[\left(\bar{b}_{j},b_{j+1}\right)+H.c.\right],
\end{eqnarray}
where we introduced $n_{\sigma}(\tau)=\bar b_{0,\sigma}(\tau)b_{0,\sigma}(\tau)$.
The coupling between the core and the bath is described by
\begin{eqnarray}
\label{eqeffectiveactioncoupling}
S_{coupling}^{\Lambda}\left(\bar{b}_{L},b_{L},\bar{b}_{L+1},b_{L+1}\right)
=-\Lambda t\ \left[\left(\bar{b}_{L},b_{L+1}\right)+H.c.\right] . 
\end{eqnarray}
\end{widetext}
We introduced the flow parameter $\Lambda$. The original model
(\ref{eqhamiltonsiam}) corresponds to $\Lambda = 1$ and for $\Lambda=0$ core
and bath are decoupled. 
In the case $L=0$, $S_{core}$ is just given by the first line of
Eq. (\ref{eqeffectiveactioncore}) and one has to replace $\Lambda t$ by
$\Lambda v$ in Eq.~(\ref{eqeffectiveactioncoupling}).

The generating functional for the connected Green's functions of the core-problem is given by 
\begin{widetext}
\begin{eqnarray}
&&\mathcal{W}_{core}\left(\bar{J},J\right)=\ln\left[\frac{1}{\mathcal{Z}_{core}}\int\mathcal{D}\left[\bar{c},c\right]
\exp\left[-S_{core}\left(\bar{c},c\right)+\sum_{i=0}^{L}\left(\bar{J}_i,c_i\right)+H.c.\right]\right]
\end{eqnarray}
with the core-fields $c = \left(b_{0},b_{1},b_{2},...,b_{L}\right)$.
$\mathcal{Z}_{core}$ is the partition function of the core-problem given by
\begin{equation}
\mathcal{Z}_{core}=\int\mathcal{D}\left[\bar{c},c\right]\ \exp\left[-S_{core}\left(\bar{c},c\right)\right].
\end{equation}
$\mathcal{W}_{core}$ can be expanded in the fields
\begin{eqnarray}
\nonumber
\mathcal{W}_{core}\left(\bar{J},J\right)=\sum_{n=0}^{\infty}\frac{(-1)^n}{n!^2}\sum_{i_1,...,i_n \atop i_1',...,i_n'}\int_0^{\beta}d\tau_{1}...\int_0^{\beta}d\tau_n \int_0^{\beta}d\tau_1'...\int_0^{\beta}d\tau_n'
&&\mathcal{G}_{core}^{c,(n)}\left(i_{1},\tau_1;...;i_{n},\tau_n|i_{1}',\tau_1';...;i_{n}',\tau_n'\right)\\
&&\times\bar{J}_{i_1}(\tau_1)...\bar{J}_{i_n}(\tau_n)J_{i_n'}(\tau_n')...J_{i_1'}(\tau_1') \, , 
\end{eqnarray}
such that
\begin{eqnarray}
\mathcal{G}_{core}^{c,(n)}\left(i_{1},\tau_1;...;i_{n},\tau_n|i_{1}',\tau_1';...;i_{n}',\tau_n'\right)
= \left. \frac{\delta^{2n}\mathcal{W}_{core}\left(\bar{J},J\right)}{\delta\bar{J}_{i_1}(\tau_1)...\delta\bar{J}_{i_n}(\tau_n)\delta J_{i_n'}(\tau_n')...\delta J_{i_1'}(\tau_1')}\right|_{J=\bar{J}=0} \, . 
\end{eqnarray}
\end{widetext}
With the definition of the $(L+1)$-component field 
$\chi~=~t(0,0,...,b_{L+1})$
we can rewrite the action (\ref{eqmicroscopicaction}) as
\begin{eqnarray}
\nonumber
S &=& S_{core}\left(\bar{c},c\right)-\sum_{i=0}^{L}\left(\bar{c_i},\Lambda\chi_i\right)-H.c.\\
&&+S_{bath}\left(\bar{b}_{L+1},b_{L+1},\bar{b}_{L+2},b_{L+2},...\right) \, . 
\end{eqnarray}
Now we can formally integrate out the $c$-fields, 
\begin{eqnarray*}
&&
\int\mathcal{D}\left[\bar{c},c\right]\exp\left[-S_{core}\left(\bar{c},c\right)+\sum_{i=0}^{L}\left(\bar{c}_i,\Lambda\chi_i\right)+H.c.\right]\\
&&=\mathcal{Z}_{core}\exp\left[\mathcal{W}_{core}\left(\Lambda\bar{\chi},\Lambda\chi\right)\right] ,
\end{eqnarray*}
yielding an effective action
for the bath
\begin{widetext}
\begin{eqnarray}
\nonumber
\mathcal{Z}&=&\int\mathcal{D}\left[\bar{c},c,\bar{b}_{L+1},b_{L+1},...\right]
\exp\left[-S_{core}\left(\bar{c},c\right)+\sum_{i=0}^{L}\left(\bar{c}_i,\Lambda\chi_i\right)+H.c.
-S_{bath}\left(\bar{b}_{L+1},b_{L+1},\bar{b}_{L+2},b_{L+2},...\right)\right]\\
\nonumber
&=&\int\mathcal{D}\left[\bar{b}_{L+1},b_{L+1},\bar{b}_{L+2},b_{L+2},...\right]\exp\left[-S^{eff}_{bath}\left(\bar{b}_{L+1},b_{L+1},\bar{b}_{L+2},b_{L+2},...\right)\right]
.
\end{eqnarray}
When we expand $\mathcal{W}_{core}\left(\Lambda\bar{\chi},\Lambda\chi\right)$
in the $\bar{\chi},\chi$-fields the effective action has the following form  
\begin{eqnarray}
\nonumber
&&S^{eff}_{bath}\left(\bar{b}_{L+1},b_{L+1},\bar{b}_{L+2},b_{L+2},...\right)\\
\nonumber
&=&S_{bath}\left(\bar{b}_{L+1},b_{L+1},\bar{b}_{L+2},b_{L+2},...\right)
-\mathcal{W}_{core}\left(\Lambda\bar{\chi},\Lambda\chi\right)\\
\nonumber
&=&S_{bath}\left(\bar{b}_{L+1},b_{L+1},\bar{b}_{L+2},b_{L+2},...\right)
-\sum_{n=0}^{\infty}\frac{(-1)^n\Lambda^{2n}}{n!^2}\sum_{i_1,...,i_n \atop i_1',...,i_n'}\int_0^{\beta}d\tau_{1}...\int_0^{\beta}d\tau_n \int_0^{\beta}d\tau_1'...\int_0^{\beta}d\tau_n'\\
\nonumber
&&\times\mathcal{G}^{c,(n)}_{core}\left(i_1,\tau_1;...;i_n,\tau_n|i_1',\tau_1;,...;i_n',\tau_n'\right)
\bar{\chi}_{i_1}(\tau_1)...\bar{\chi}_{i_n}(\tau_n)\chi_{i_n'}(\tau_n')...\chi_{i_1'}(\tau_1') \, . 
\end{eqnarray}
\end{widetext}
In the following we neglect the term with $n=0$, which does not contain any
fields. Furthermore,
we truncate the sum over $n$ after $n=2$. This means we consider only the first
and second order of the expansion and neglect all 
correlation-functions $\mathcal{G}^{c,(\geq3)}_{core}$. When we transform the action to Matsubara frequencies we get
\begin{widetext}
\begin{eqnarray}
\nonumber
&& S^{eff}_{bath}\left(\bar{b}_{L+1},b_{L+1},\bar{b}_{L+2},b_{L+2},...\right)\\
\nonumber
&=&S_{bath}\left(\bar{b}_{L+1},b_{L+1},\bar{b}_{L+2},b_{L+2},...\right)
+\frac{\Lambda^2}{\beta}\sum_{i\omega}\sum_{i_1,i_1'}\bar{\chi}_{i_1}(i\omega)\mathcal{G}^{c,(1)}_{core}\left(i\omega,i_1,i_1'\right)\chi_{i_1'}(i\omega)\\
\nonumber
&&-\frac{\Lambda^4}{4\beta^3}\sum_{i\omega_1,i\omega_2,\atop i\omega_1',i\omega_2'}\sum_{i_1,i_2,\atop i_1',i_2'}\bar{\chi}_{i_1}(i\omega_1)\bar{\chi}_{i_2}(i\omega_2)
\mathcal{G}^{c,(2)}_{core}\left(i\omega_1,i_1;i\omega_2,i_2|i\omega_1',i_1';i\omega_2',i_2'\right)
\chi_{i_1'}(i\omega_1')\chi_{i_2'}(i\omega_2')\delta_{\omega_1+\omega_2,\omega_1'+\omega_2'}\\
\nonumber
&=&S_{bath}\left(\bar{b}_{L+1},b_{L+1},\bar{b}_{L+2},b_{L+2},...\right)
+\frac{(\Lambda t)^2}{\beta}\sum_{i\omega}\sum_{\sigma}\bar{b}_{L+1,\sigma}(i\omega)\mathcal{G}^{c,(1)}_{core,\sigma}\left(i\omega,b_L,b_L\right)b_{L+1,\sigma}(i\omega)\\
\nonumber
&&-\frac{(\Lambda t)^4}{4\beta^3}\sum_{i\omega_1,i\omega_2,\atop i\omega_1',i\omega_2'}\sum_{\sigma_1,\sigma_2,\atop \sigma_1',\sigma_2'}\bar{b}_{L+1,\sigma_1}(i\omega_1)\bar{b}_{L+1,\sigma_2}(i\omega_2)
\mathcal{G}^{c,(2)}_{core}\left(i\omega_1,b_{L},\sigma_1;i\omega_2,b_{L},\sigma_2|i\omega_1',b_{L},\sigma_1';i\omega_2',b_{L},\sigma_2'\right)\\
\nonumber
&&\times b_{L+1,\sigma_1'}(i\omega_1')b_{L+1,\sigma_2'}(i\omega_2')\delta_{\omega_1+\omega_2,\omega_1'+\omega_2'} \delta_{\sigma_1+\sigma_2,\sigma_1'+\sigma_2'} \, . 
\end{eqnarray}
In the effective-bath theory there is a local interaction on bath site
$L+1$. The other bath sites ($L+2$,$L+3$,...) remain noninteracting and can be
integrated out. This leads to the local effective action
\begin{eqnarray} \nonumber 
S^{eff}_{bath}\left(\bar{b}_{L+1},b_{L+1}\right)
&=&-\frac{1}{\beta}\sum_{i\omega}\sum_{\sigma}\bar{b}_{L+1,\sigma}(i\omega)
\left(i\omega-(\Lambda
  t)^2\mathcal{G}^{c,(1)}_{core,\sigma}\left(i\omega,b_L,b_L\right)-t^2
  g_b\left(i\omega,b_{1},b_{1}\right)\right)b_{L+1,\sigma,i\omega}\\ 
\nonumber
&&-\frac{(\Lambda t)^4}{4\beta^3}\sum_{i\omega_1,i\omega_2,\atop i\omega_1',i\omega_2'}\sum_{\sigma_1,\sigma_2,\atop \sigma_1',\sigma_2'}\bar{b}_{L+1,\sigma_1}(i\omega_1)\bar{b}_{L+1,\sigma_2}(i\omega_2)
\mathcal{G}^{c,(2)}_{core}\left(i\omega_1,b_{L},\sigma_1;i\omega_2,b_{L},\sigma_2|i\omega_1',b_{L},\sigma_1';i\omega_2',b_{L},\sigma_2'\right)\\ 
&& b_{L+1,\sigma_1'}(i\omega_1')b_{L+1,\sigma_2'}(i\omega_2') \delta_{\omega_1+\omega_2,\omega_1'+\omega_2'}\delta_{\sigma_1+\sigma_2,\sigma_1'+\sigma_2'} \,.  \label{eqeffectiveaction}
\end{eqnarray}
\end{widetext}
The local correlation functions of the core, $\mathcal{G}^{c,(1)}_{core}$ and
$\mathcal{G}^{c,(2)}_{core}$  can be calculated from the Lehmann representation,
which is given in appendix \ref{sec:appendixcore}. Note that for $\Lambda =0$, the bath theory is
noninteracting. The exact solution of this serves as an initial condition for
the fRG flow in $\Lambda$.

\subsection{Relation to the Dot self-energy}
\label{sec:effbath2}
In the effective theory (\ref{eqeffectiveaction}) the bath site $L+1$ is
now interacting with a frequency dependent term, while in the original theory
(\ref{eqmicroscopicaction}) it was noninteracting. The self-energy and all
higher irreducible vertex-functions are local on the
dot-site by construction . Nevertheless the local Green's function  of the
coupled problem on bathsite $L+1$ is nontrivial and depends on the
dot self-energy, as can be seen  in
(\ref{eqgreensfunctionfirstbathsite}) or
(\ref{eqgreensfunctionsndbathsite}). This Green's function can also be derived
in the setup of the effective theory and one can use the identity
$\mathcal{G}_{\sigma}\left(i\omega_{n},b_{L+1},b_{L+1}\right)=\mathcal{G}^{eff}_{\sigma}\left(i\omega_{n},b_{L+1},b_{L+1}\right)$
with
\begin{eqnarray}
&&\mathcal{G}^{eff}_{\sigma}\left(i\omega_{n},b_{L+1},b_{L+1}\right) =\\
&&\left[i\omega-(\Lambda t)^2\mathcal{G}^{c,(1)}_{core,\sigma}\left(i\omega,b_L,b_L\right)
-t^2
g_b\left(i\omega\right)-\Sigma_{b,\sigma}(i\omega)\right]^{-1} \nonumber
\end{eqnarray}
to get a relation between the dot self-energy and the effective self-energy $\Sigma_b$ on
bath site $L+1$. These relations depend on $L$ and for $L=0,1,2,3$ we get the
relations given in Eq. (\ref{eqrelationdotbathl0})-(\ref{eqrelationdotbathl3})
in appendix \ref{sec:appendixrelation}.

\section{Functional RG flow equations}
\label{sec:FRGeqs}
The effective action in Eq.~(\ref{eqeffectiveaction}) reduces to a
noninteracting model for $\Lambda=0$, because the interaction term is
proportional to $\Lambda^4$. This represents a simple starting point for a fRG
flow in $\Lambda$. In order to use the fRG formalism for one-particle
irreducible (1PI) vertices \cite{Wet93,Sal01,Met12}, the flow parameter
$\Lambda$ should only occur in the quadratic part of the action. This can be
achieved for any $\Lambda \not=0$ by  
rescaling the fields $b_{L+1}\to b_{L+1}/\Lambda$ and $\bar{b}_{L+1}\to
\bar{b}_{L+1}/\Lambda$. This leads to the quadratic part of the effective action,
\begin{widetext}
\begin{eqnarray}\label{eqeffectiveactionumparametrisiert}
\nonumber
S^{eff,0}_{bath}\left(\bar{b}_{L+1},b_{L+1}\right)
&=&-\frac{1}{\beta}\sum_{i\omega}\sum_{\sigma}\bar{b}_{L+1,\sigma}(i\omega)
Q^{\Lambda}_{\sigma}\left(i\omega\right)b_{L+1,\sigma,i\omega} .
\end{eqnarray}
\end{widetext}
with
\begin{equation}
Q^{\Lambda}_{\sigma}\left(i\omega\right)=\frac{i\omega}{\Lambda^2}-t^2\mathcal{G}^{c,(1)}_{core,\sigma}\left(i\omega,b_L,b_L\right)-\frac{t^2}{\Lambda^2}
  g_b\left(b_{1},b_{1},i\omega\right)
\end{equation}
and quartic part which does not depend on $\Lambda$ anymore. 
Note that the rescaling changes correlation functions of different order in
the fields differently, but in the end we will study the case $\Lambda =1$.  
The one-particle-irreducible-(1PI)-vertex-functions on scale $\Lambda$ can be
calculated by an infinite set of exact 
flow-equations. \cite{Wet93,Sal01,Met12} If we neglect the flow of the
three-particle-vertex and of all higher vertex functions we get a closed set of
equations for the self-energy 
$\Sigma_{b}^{\Lambda}$ and the two-particle vertex-function
$\Gamma_{b}^{\Lambda}$,
\begin{eqnarray}
\label{eqflussgleichungselbstenergie}
&&\frac{d}{d\Lambda}\Sigma_{b}^{\Lambda}(k';k) =
-\textrm{Tr}\left[S^{\Lambda}\Gamma_{b}^{\Lambda}(k',.;k,.)\right]\\
\label{eqflussgleichungvertex}
&&\frac{d}{d\Lambda}\Gamma_{b}^{\Lambda}(k_{1}',k_{2}';k_{1},k_{2}) =
\textrm{Tr}\left[
S^{\Lambda}\Gamma_{b}^{\Lambda}(k_{1}',k_{2}',.;k_{1},k_{2},.)
\right] \nonumber\\
&&-\textrm{Tr}\left[S^{\Lambda}\Gamma_{b}^{\Lambda}(.,.;k_{1},k_{2})[\mathcal{G}^{\Lambda}]^{T}\Gamma_{b}^{\Lambda}(k_{1}',k_{2}';.,.)\right]\nonumber
\\
&&-\textrm{Tr}\left[S^{\Lambda}\Gamma_{b}^{\Lambda}(k_{1}',.;k_{1},.)\mathcal{G}^{\Lambda}\Gamma_{b}^{\Lambda}(k_{2}',.;k_{2},.)\right]\nonumber\\
&&-[k_{1}'\leftrightarrow k_{2}'] - [k_{1}\leftrightarrow k_{2}]
+[k_{1}'\leftrightarrow k_{2}', k_{1}\leftrightarrow k_{2}],
\end{eqnarray}
in which $\mathcal{G}^{\Lambda}$ is the full propagator and
$S^{\Lambda}$ is the so called single-scale propagator defined by
\begin{equation}
S^{\Lambda}=\mathcal{G}^{\Lambda}\frac{d}{d\Lambda}\left[Q^{\Lambda}\right]\mathcal{G}^{\Lambda}.
\end{equation}
The $k_{i}^{(')}$ denote one-particle-quantum-numbers (in our case
Matsubara-freqency, spin and site-index) and the trace is defined with respect
to these quantum numbers. 
In the case $B=0$, the flow-equations (\ref{eqflussgleichungselbstenergie})
and (\ref{eqflussgleichungvertex}) can be further simplified by using the
spin-rotation-invariance of the effective action
(\ref{eqeffectiveactionumparametrisiert}), which is described in 
appendix \ref{sec:appendixfrg}. 

By integrating the flow-equations from $\Lambda=0$ to $\Lambda=1$ we can
derive the self-energy of the effective bath theory. Using the relations
(\ref{eqrelationdotbathl0}) -  
(\ref{eqrelationdotbathl3}) we obtain the dot self-energy $\Sigma_d\left(i\omega\right)$.

In the simplest approximation one neglects the flow of the two-particle-vertex and  
integrates equation (\ref{eqflussgleichungselbstenergie}) with
$\Gamma_b^{\Lambda}=\Gamma_b^{\Lambda=0}$ (in the following called 'approximation
1'), where $\Gamma_b^{\Lambda=0}$ is given by Eqs. (\ref{eqrelationgammakopplung}) and (\ref{eqanfangsbedingungvertex}). 
If one integrates the full set of Eqs. (\ref{eqflussgleichungselbstenergie})
and (\ref{eqflussgleichungvertex}) (called 'approximation 2') the
numerical effort scales with the third power of the number of
Matsubara-frequencies. 

Motivated by the fullfillment of Ward-identities in the RG-flow, the following
replacement in the flow-equation for the vertex-function was
proposed,\cite{Kat04} 
\begin{equation}
S^{\Lambda}\rightarrow
-\frac{d\mathcal{G}^{\Lambda}}{d\Lambda}=S^{\Lambda}-\mathcal{G}^{\Lambda}\frac{d\Sigma_{b}^{\Lambda}}{d\Lambda}\mathcal{G}^{\Lambda} 
\, .  
\end{equation}
This replacement is used in all following calculations.

Instead of doing the fRG-flow in the effective bath-theory
(\ref{eqeffectiveactionumparametrisiert}) one can also derive flow-equations
for  the dot self-energy (\ref{eqrelationdotbathl0})-
(\ref{eqrelationdotbathl3}) and the two-particle-(1PI)-vertex on the dot-site
with the core-(1PI)-vertex-functions 
as initial condition. For $L>0$ these flow-equations have a more complicated structure than
in our case and the calculations are easier in the setup of the 
effective bath-theory. For $L=0$ and approximation 1 we compared both
schemes. It turned out that neglecting the flow of the 2-particle-vertex 
of the effective bath-theory leads to better results than doing an analogous
approximation for the 2-particle-vertex on the dot. 

\section{Numerical results}
\label{sec:results}

In the following we present our results for different sizes of the core ($L =
0, 1$ and $L=3$). The data shown here was produced by integrating
Eqs. (\ref{eqflussgleichungselbstenergie}, \ref{eqflussgleichungvertex})
numerically from $\Lambda=0$ to $\Lambda =1$, using typically 100 to 200
Matsubara frequencies at temperatures varying between $\beta=20/\Delta_0$ and
$\beta=50/\Delta_0$. We set $\Delta_0=1$ giving the energy scale and in most
cases consider particle-hole symmetry, $\epsilon_d = -U/2$.
The results are compared with other fRG approaches and benchmark NRG
calculations. The latter are carried out with the same semi-elliptic density
of states. 
We focus on quantities that can directly be calculated from the data on the
imaginary frequency-axis. 
One of these quantities is the effective mass $m^{*}$ given by
\begin{equation}
m^{*}=z^{-1}=1-\left. \frac{d\text{Im}\Sigma_d\left(i\omega\right)}{d\omega}\right|_{\omega=0^{+}}.
\end{equation}
In the Kondo-regime the quasiparticle weight $z$ determines the width of the Kondo-resonance and
is expected to scale exponentially with the interaction strength. 
We furthermore calculate the linear conductance of the dot $G = \sum_{\sigma}
G_{\sigma}$ given by (in the following we set $\hbar=e^2=1$) 
\begin{eqnarray}\label{eqleitfaehigkeit}
\nonumber
G_{\sigma} &=& \frac{1}{2}\pi v^2\int d\omega A_{d,\sigma}\left(\omega\right)
\rho_b\left(\omega\right) \left(-\frac{\partial
    n_F\left(\omega\right)}{\partial\omega}\right)\\ 
&\simeq& \frac{1}{2}\Delta_0 \int d\omega A_{d,\sigma}\left(\omega\right)
\left(-\frac{\partial n_F\left(\omega\right)}{\partial\omega}\right) \,  .  
\end{eqnarray}
In the second line we used that the derivative of the Fermi-function is
sharply peaked at low temperature at $\omega = 0$ so that the $\omega$-dependence of
$\rho_b(\omega)$ can be neglected. 
Of course the conductance is derived by an integral over the real frequency
axis and at first sight one has also to perform an analytic continuation. 
To circumvent this, we follow an approach, proposed in
Ref. \onlinecite{Kar10}, that does not require an analytic continuation. In
this approach $G$ follows from the formula 
\begin{equation}
\label{eqleitfaehigkeit2}
G_{\sigma} \simeq \Delta_0 T \sum_{\alpha>0} R_{\alpha}
\text{Im}\frac{d\mathcal{G}_{\sigma}(i\tilde{\omega}_{\alpha})}{d\tilde{\omega}_{\alpha}},  
\end{equation}
where the imaginary frequencies $i\tilde{\omega}_{\alpha}$ and the weights
$R_{\alpha}$ are defined in Ref. \onlinecite{Kar10}. The frequencies
$i\tilde{\omega}_{\alpha}$ differ from the original Matsubara-frequencies and
we determine
$\frac{d\mathcal{G}_{\sigma}(i\tilde{\omega}_{\alpha})}{d\tilde{\omega}_{\alpha}}$
from a Pade-approximation. 

We have also done the analytic continuation
$\mathcal{G}\left(i\omega_n,d,d\right)\rightarrow\mathcal{G}\left(\omega+i 
  0^{+},d,d\right)$ to the real frequency axis using a
Pade-algorithm described in Ref. \onlinecite{Vid77}. 
As the analytical continuation of numerical data is mathematically an
ill-defined problem we did not obtain numerically stable and meaningful results for all parameter sets.

\subsection{The case $L=0$}
\label{sec:resultsL0}
For $L=0$ the 'core' is given by an isolated dot site. In this case the
core-correlation-functions can be calculated analytically \cite{Haf09} and one
can also derive analytical results in the setup of the effective bath-theory.
\cite{Joe10} 

\begin{figure}[htbp]
    \centering
      \includegraphics[width=0.45\textwidth]{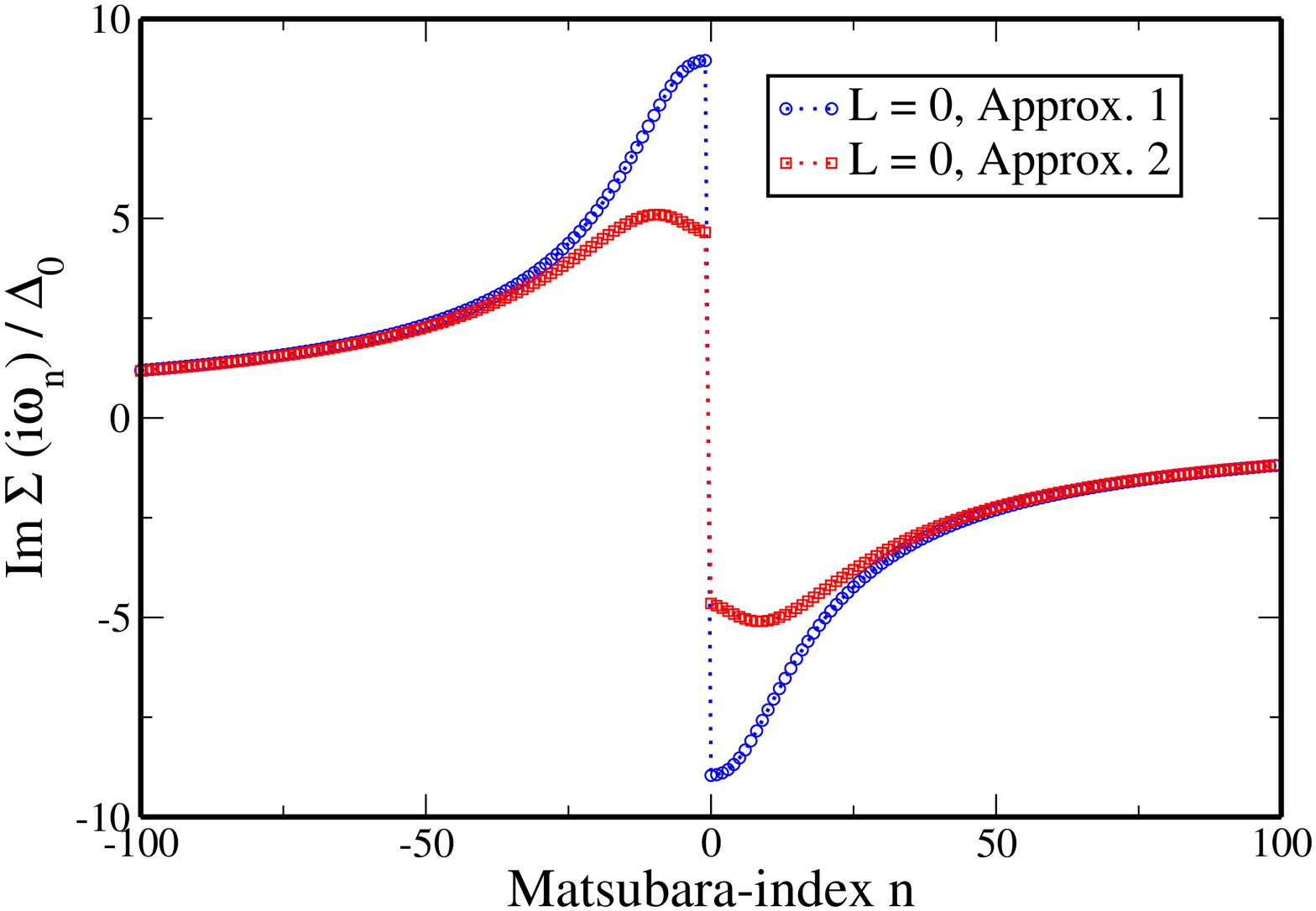}
      \includegraphics[width=0.45\textwidth]{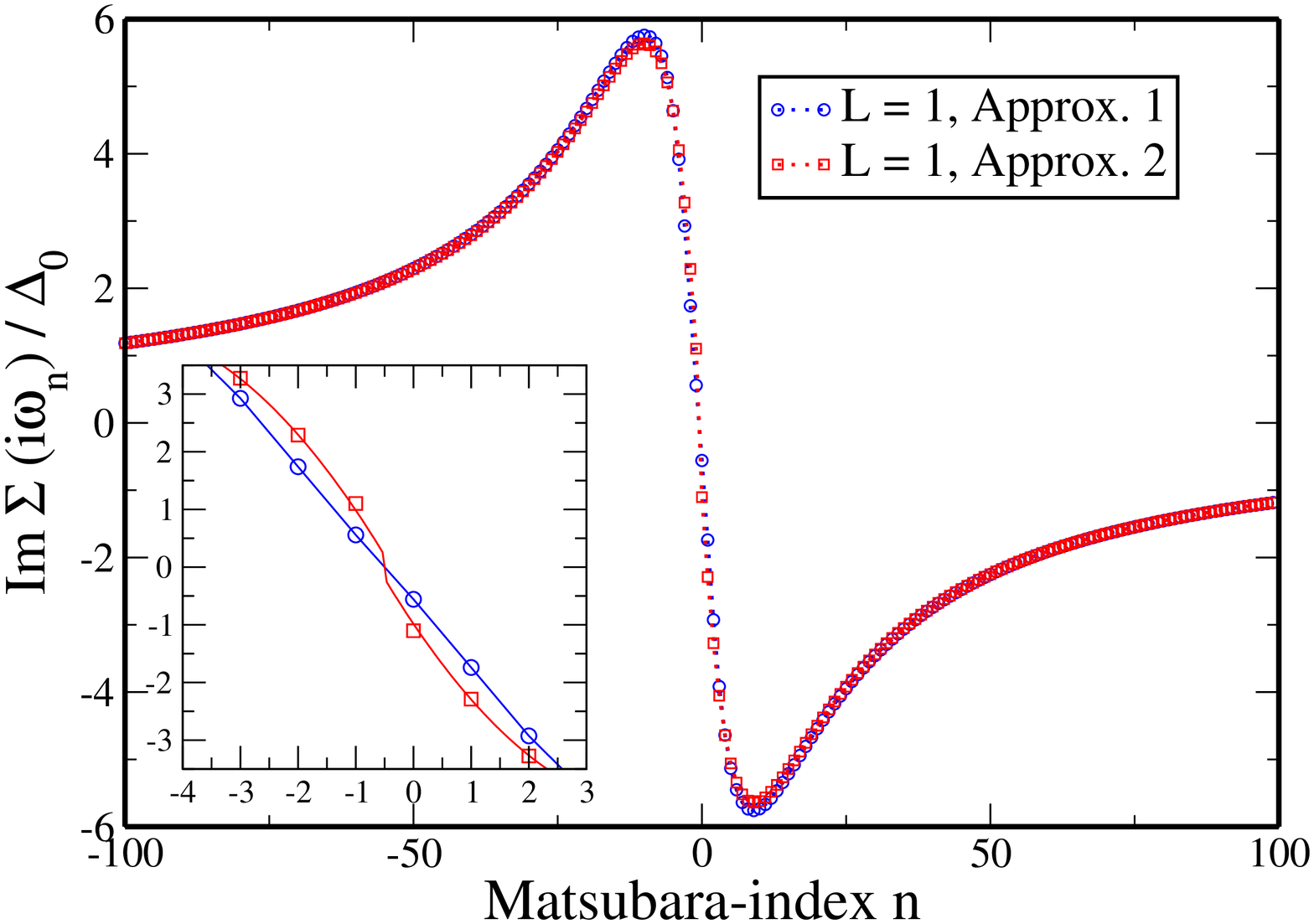}
    \caption[vglselbstenergien]{(color online) Comparison of approximations 1
      and 2 to the Matsubara-self-energy for $U = 10\Delta_0$, $\beta = 30 /
      \Delta_0$ and $L = 0, 1$.  For $L=0$ one gets
      a discontinuity of $\text{Im}\Sigma(i\omega_n)$ between positive and
      negative Matsubara-frequencies, which is reduced if one increases the
      level of approximation. 	For $L=1$, approximation 1 the self-energy is
      continuous for small frequencies. As shown in the inset one obtains a
      small step between positive and negative frequencies 	in approximation 2.} 
    \label{pictvglmatsubaraselbstenergien}
\end{figure}

The initial self-energy at $\Lambda=0$ for $\epsilon_d = -U/2$  is given by
\begin{equation}
\Sigma_d\left(i\omega\right)=\frac{U}{2}+\frac{U^2}{4 i\omega},
\end{equation}
which is the atomic limit result. $\Sigma_d\left(i\omega\right)$
diverges at $i\omega=0$.  
In Fig.~\ref{pictvglmatsubaraselbstenergien} (upper panel) we show the
$i\omega$-dependenc of the self-energy at the end of the flow for $\Lambda=1$
on the dot for $U=10\Delta_0$,  $\beta=30/\Delta_0$ and for the particle-hole
symmetric case, $V_g=0$,  computed with the described fRG-flow in both
approximations 1 and 2. In the calculation we included 200
Matsubara-frequencies.  The divergence of the self-energy at $i\omega=0$ has
disappeared, but there is still a discontinuity, which is not cured by the
flow. This shows the flow equations are not able to restore the expected local Fermi liquid
properties of the SIAM, if we start with the atomic solution. The height of
the unphysical discontinuity becomes however smaller in approximation 2  
compared to approximation 1. 

In the spectral density derived from a Pade-approximation to our numerical
data at half filling (not shown) two slightly broadened atomic limit peaks at $\pm U/2$, 
but no central Kondo-resonance at small frequencies appears. Hence the $L=0$ 
approximation fails to describe the screening of the local spin-1/2-moment by
the conduction-electrons. 
This screening and singlet formation should develop when $\Lambda$ is switched
on, 
while at $\Lambda=0$ the local moment is unscreened. 
We assume that this strong mismatch is the reason for the
non-occurence of the Kondo-resonance in this approximation. 

Our results are consistent with the findings in Ref. \onlinecite{Haf09}, where
a superperturbation approach to the Anderson model is developed. In this
approach a finite local cluster containing the correlated dot-site is solved
exactly. The correlation-functions of this cluster then serve as input for an
effective theory of dual-fermion-fields. Like in our setup, no Kondo-resonance
is found when the cluster contains only the correlated bath-site.

\subsection{The cases $ L=1, 2$ and $3$}

In the case $L=1$ the isolated core consists of an interacting site coupled by a
hopping-term $v$ to a noninteracting bath-site. This model is still analytically
solvable. The ground-state at half-filling is a spin-singlet-state. For
$\epsilon_d=-U/2$  and in the limit $v\ll U$ and this state is given by 
\begin{equation}
\label{eqgroundstatecoreL1}
| S=0 \rangle =
\frac{4v}{U}\left(\ket{\uparrow\downarrow,e}+\ket{e,\uparrow\downarrow}\right)-\left(1-\frac{8
    v^2}{U^2}\right)\left(\ket{\uparrow,\downarrow}-\ket{\downarrow,\uparrow}\right) 
\end{equation}
with energy $\frac{1}{4}\left(-U-\sqrt{U^2+64 v^2}\right)\overset{v\ll
  U}{\approx} -\frac{U}{2}-\frac{8 v^2}{U}$. The first entry in $|. ,
. \rangle$ is the correlated site, the second the additional uncorrelated core
site. $e$ stands for an empty site. Now, in contrast to the case $L=0$, the
local-moment on the dot is already in a singlet state for $\Lambda=0$. 
This is a much better starting point to describe features of the Kondo effect. 
As can be seen in Fig.~\ref{pictvglmatsubaraselbstenergien} the self-energy at the end
of the flow for $L=1$ is continuous at $i\omega=0$.
For $L=2$ the isolated core self-energy has a similar shape as in the
$L=0$-case and for $\Lambda=1$ we get a finite step between positive and
negative Matsubara-frequencies. In its groundstate the core carries 
again a finite $s=1/2$-moment, doublet ground state,  in this case which does
not become screened when we switch on the coupling to the bath in the
fRG-flow. This shows once more the importance of choosing a 
core with spin-singlet groundstate for an at least qualitatively correct
description of Kondo screening in this setup. The next larger core size with a
singlet groundstate contains $L=3$ bath sites. Numerically the calculation of
the two-particle vertex function is limited due to the exponential growth of
the core-Hilbert-space. We just used approximation 1 in the $L=3$ case,
because here we only need to calculate the vertex for two 
instead of three independent frequencies. 

Let us now discuss the numerical results in more detail. The spectrum of the
isolated two-site core with $L=1$ consists of four delta-peaks.  Two of them
are located at 
$\epsilon_{1,2}=\pm \frac{1}{4}\left(\sqrt{U^2+64 v^2}+\sqrt{U^2+16
    v^2}\right)\overset{v\ll U}{\approx}\pm\left(\frac{U}{2}+\frac{10
    v^2}{U}\right)$, which belong to 
excitations from the ground-state (\ref{eqgroundstatecoreL1}) to the
one-particle-state $\frac{2v}{U}\ket{\sigma,e}+\left(1-\frac{2
    v^2}{U^2}\right)\ket{e,\sigma}\quad (v\ll U)$ and its corresponding
three-particle-state, which is connected  
by a particle-hole-transformation. In the limit $v\rightarrow 0$ they are
equal to the atomic $\pm U/2$-excitations of the L=0-core. When we switch on
the coupling to the bath in the fRG-flow, they evolve into hybridization
broadened peaks. The other two peaks in the spectrum of the L=1-core lie at
$\epsilon_{3,4}=\pm \frac{1}{4}\left(\sqrt{U^2+64 v^2}-\sqrt{U^2+16
    v^2}\right)\overset{v\ll U}{\approx} 
\pm\frac{6 v^2}{U}$. They belong to excitations from the ground-state to the
one-particle-state $\frac{2v}{U}\ket{\sigma,e}+\left(1-\frac{2
    v^2}{U^2}\right)\ket{e,\sigma}\quad (v\ll U)$ and its corresponding 
three particle-state. Note, that due to these excitations, there is a finite
spectral weight near zero-energy  already for $\Lambda=0$ that can evolve
into a central Kondo resonance. 

\begin{figure}[htbp]
    \centering
    \includegraphics[width=0.50\textwidth]{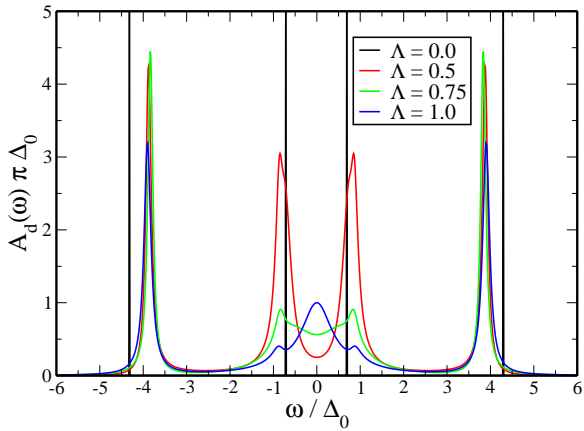}
    \caption[evolutionofspectra]{(color online) $L=1$ dot-spectra for several
      values of $\Lambda$ at half filling and $U=6\Delta_0$ and
      $\beta=50/\Delta_0$, obtained from a Pade-approximation to our numerical 
	  data on the imaginary frequency axis. The atomic limit peaks become
          slightly broadened and their position changes from 4.2$\Delta_0$ to 3.9$\Delta_0$. At small
          frequencies a central resonance with height
          $A_d(\omega=0)=1/\pi\Delta_0$ is emerging during the fRG flow from
          $\Lambda=0$ to $\Lambda=1$.} 
    \label{pictevolutionofspectra}
\end{figure}

It turns out that  already in the most simple approximation 1 we get a
quasiparticle resonance at $\omega=0$. The change of the spectrum for
different values of $\Lambda$ is shown in Fig.~\ref{pictevolutionofspectra} for $U=6 \Delta_0$.  
The peaks $\epsilon_{1,2}$ become slightly broadened, but their positions does
not change significantly. In the end of the flow (for $\Lambda=1$) their
maxima are not located at $\pm U/2=\pm 3\Delta_0$, the position usually expected in the
wide band limit with a purely imaginary hybridization function
$\Delta(\omega)$. However, in the present case where the bandwidth is less
than $U$ the hybridization function $\Delta(\omega)$ has a finite real part
which renormalizes this position. The position of the peaks, 
$\pm 3.9\Delta_0$ turns out to be comparable with what is found in NRG
calculations. 
The small broadening of these high-energy peaks is related to the fact that
they lie outside the bandwidth $(-2\Delta_0,2\Delta_0)$ of the bath, such that the
width is purely due to self-energy effects. 
During the flow, already for small values of $\Lambda$, the low-energy peaks
$\epsilon_{3,4}$ become broadened and a central resonance at $\omega=0$ with
$A_d(\omega=0)=\frac{1}{\pi\Delta_0}$ emerges. 
Note that at the end of the flow, for $\Lambda=1$, there are still remnants of
the peaks $\epsilon_{3,4}$, which is interpreted as an artefact of the
approximation. The same artefacts are obtained for $L=3$.
 
In Fig.~\ref{pictvglmatsubaraselbstenergien} the self-energy calculated in
approximation 1 and 2 is shown. In approximation 1 the self-energy is
continuous for small frequencies and the derivative
$\left. \frac{d\text{Im}\Sigma_d\left(i\omega\right)}{d\omega}\right|_{\omega=0^{+}}$
is negative, which leads to a reduced width $z \Delta_0$ of the resonance at small frequencies.  
As shown in the inset of Fig.~\ref{pictvglmatsubaraselbstenergien} we
obtain a small step between positive and negative Matsubara frequencies in
approximation 2. This step leads to a slight broadening of the central
resonance, which decreases with decreasing temperature. Therefore it can be
understood as a physically sensible finite-temperature effect. 

\subsection{Results for the effectve mass in comparison}
\begin{figure}[htbp]
    \centering
    \begin{minipage}[t]{8 cm}
	\includegraphics[width=1.0\textwidth]{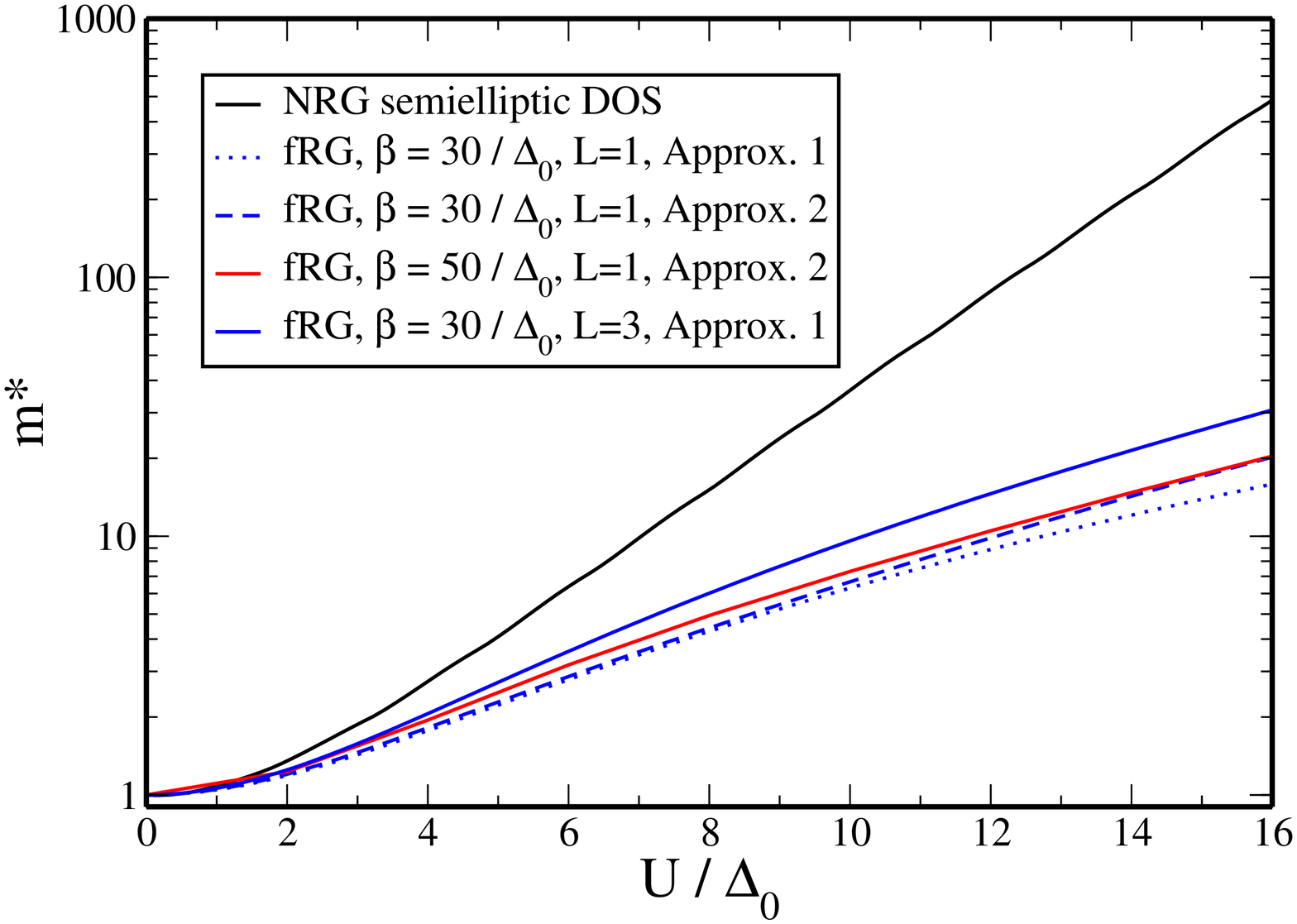}
	\caption[effecivemassL3]{(color online) Effective mass ($\beta = 30,50 / \Delta_0$) for
          L=1, approximation 1 and 2 and for L=3,
          approximation 1 in comparison with NRG-data. The NRG-data are
          calculated for a bath with semi-elliptic density of states.}
	\label{picteffectivemass1}    \end{minipage}
    \begin{minipage}[t]{8 cm}
	\includegraphics[width=1.0\textwidth]{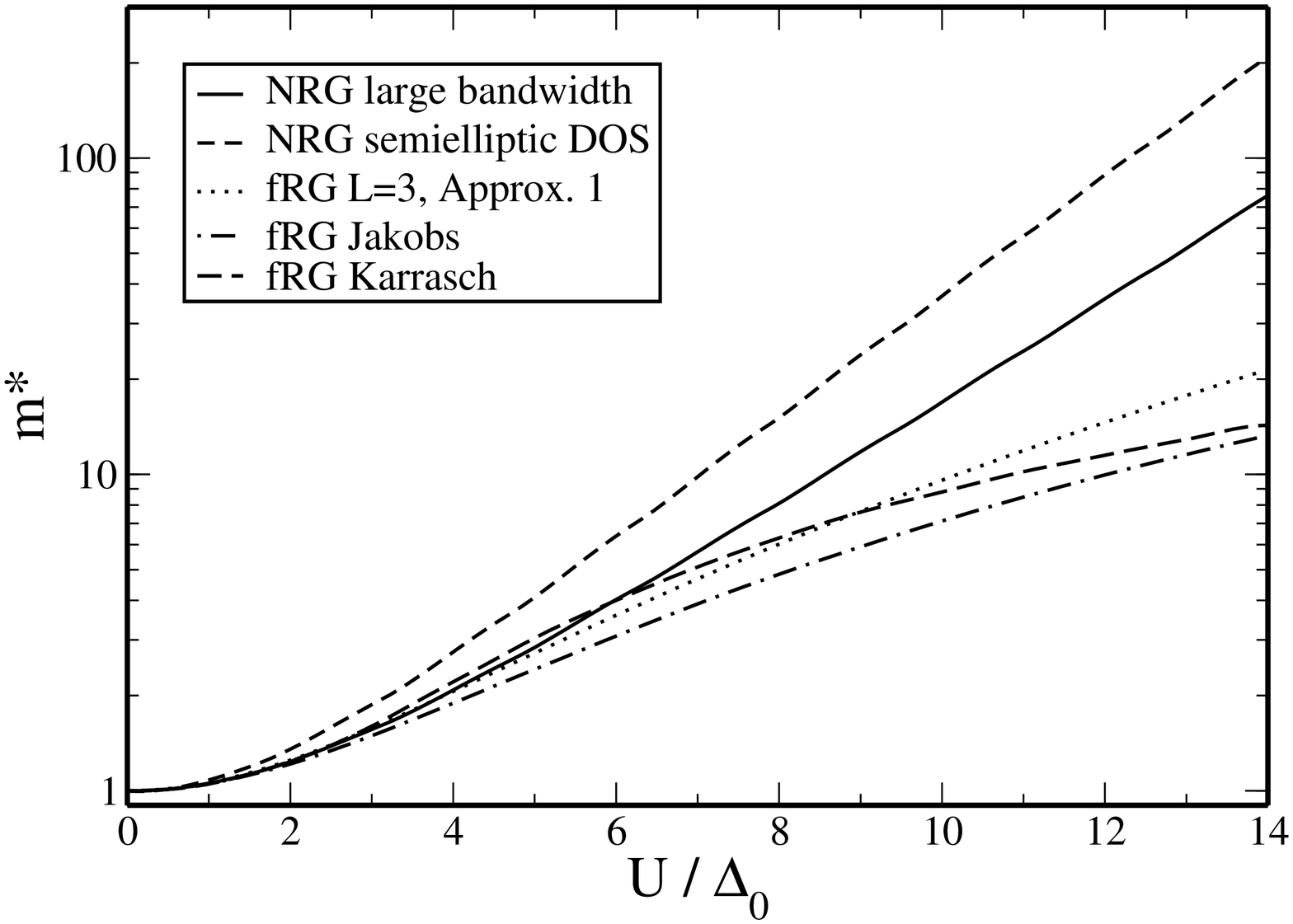}
	\caption[effecivemassL3]{Effective mass
          ($\beta=30/\Delta_0$) for L=3, approximation 1 in comparison with fRG
          data from Ref. \onlinecite{Jak10} and approximation 1 in
          Ref. \onlinecite{Kar08}. As reference data we show NRG-calculations
          for a semielliptic density of states and in the wide-band-limit.} 
	\label{picteffectivemass2}    \end{minipage}
\end{figure}

In Fig.~\ref{picteffectivemass1} we show the effective mass for $L=1$,
approximation 1 and 2 and $L=3$, approximation 1 in comparison with NRG data
as function of the interaction strength $U$. The NRG-data is calculated at
$T=0$ for a semi-elliptic bath-density of states. While the qualitative
behaviour is similar, the effective mass from the fRG-calculations is
systematically too small compared with the NRG data and we can not reproduce
the exponential Kondo-scale quantitatively. For interaction-strengths $U \sim
8-9 \Delta_0$ the Kondo-scale $T_K=W\sqrt{\frac{2\Delta_0}{\pi U}}\exp\left(-\frac{\pi U}{8\Delta_0}\right)$
becomes comparable with the temperature $T_K
\approx \frac{1}{\beta}$, which we expect to be part of the reason for the
deviations from the NRG result at large values of $U$. Note the slight increase of $m^{*}$ with decreasing temperature in
Fig. \ref{picteffectivemass1}. 
Our results for the effective mass fall in the range of other fermionic
fRG-approaches to the Anderson-model \cite{Kar08,Jak10}
(cf. Fig. \ref{picteffectivemass2}), that are calculated for a bath in the
wide-band-limit. A direct comparison of the data needs to take into account
the fact that as can be seen from the NRG-data in Fig. \ref{picteffectivemass2} the
effective mass for a semielliptic density of states with finite bandwidth is  
in general larger than for a bath in the wide-band-limit. The failure in
reproducing the exponential Kondo-scale in the effective mass precisely is
however common to all finite-frequency fRG-approaches to the Anderson model. Note, that all
fRG-approaches truncate the hierarchy of flow-equations after the four-point
level. Hence, we expect that this approximation is the reason for this
deviation. 

\subsection{Results for the conductance}
\begin{figure}[htbp]
    \centering
    \begin{minipage}[t]{8 cm}
      \includegraphics[width=1.0\textwidth]{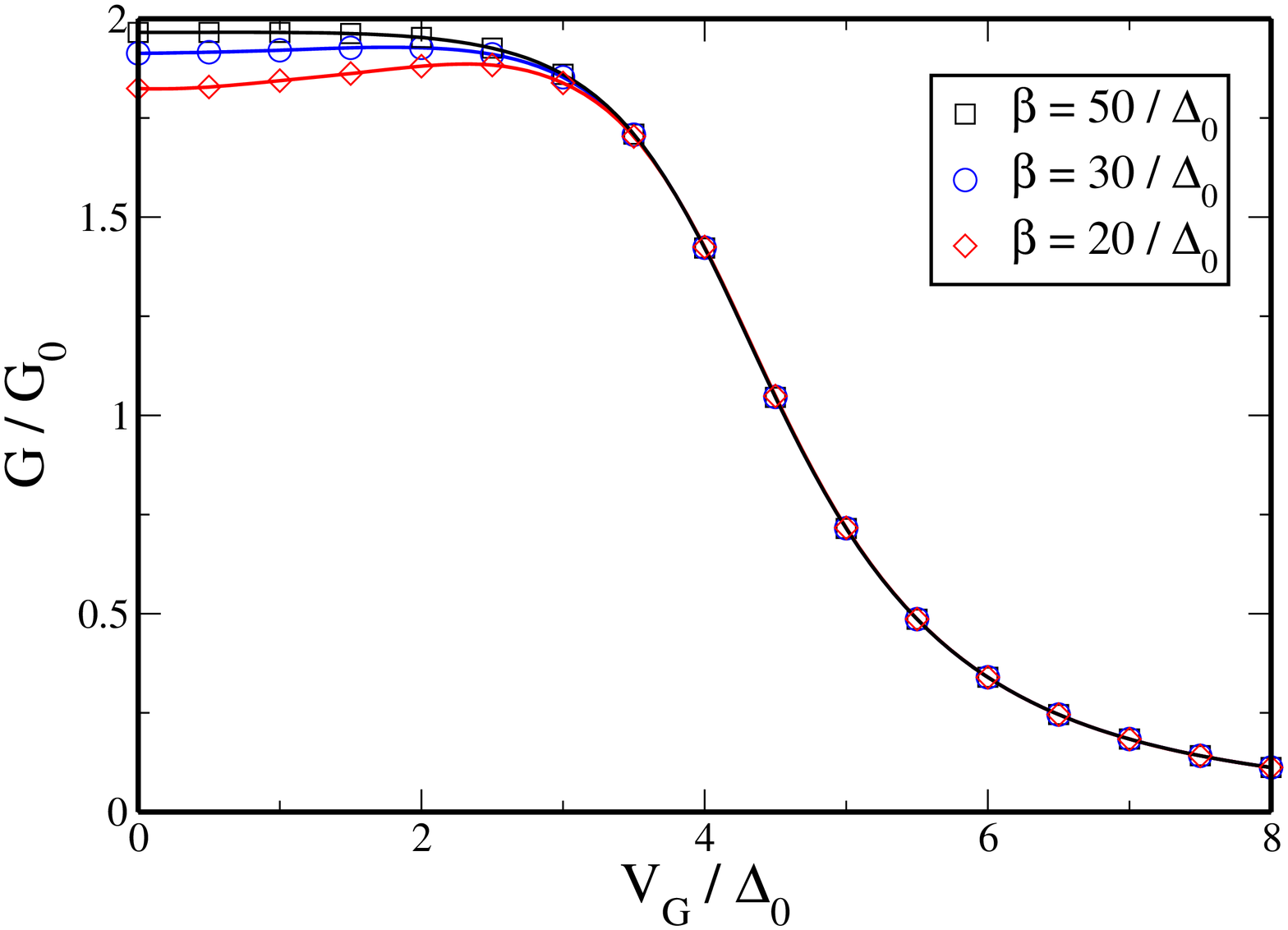}
      \caption[compleitfaehigkeittemp]{(color online) Comparison of the linear
        conductance for $U=8\Delta_0$ and $\beta=20,30,50 / \Delta_0$,
        calculated in approximation 1. The value  $G(V_G = 0)$ decreases quadratically with
        increasing temperature.} 
      \label{pictcompleitfaehigkeittemp}
    \end{minipage}
    \begin{minipage}[t]{8 cm}
      \includegraphics[width=1.0\textwidth]{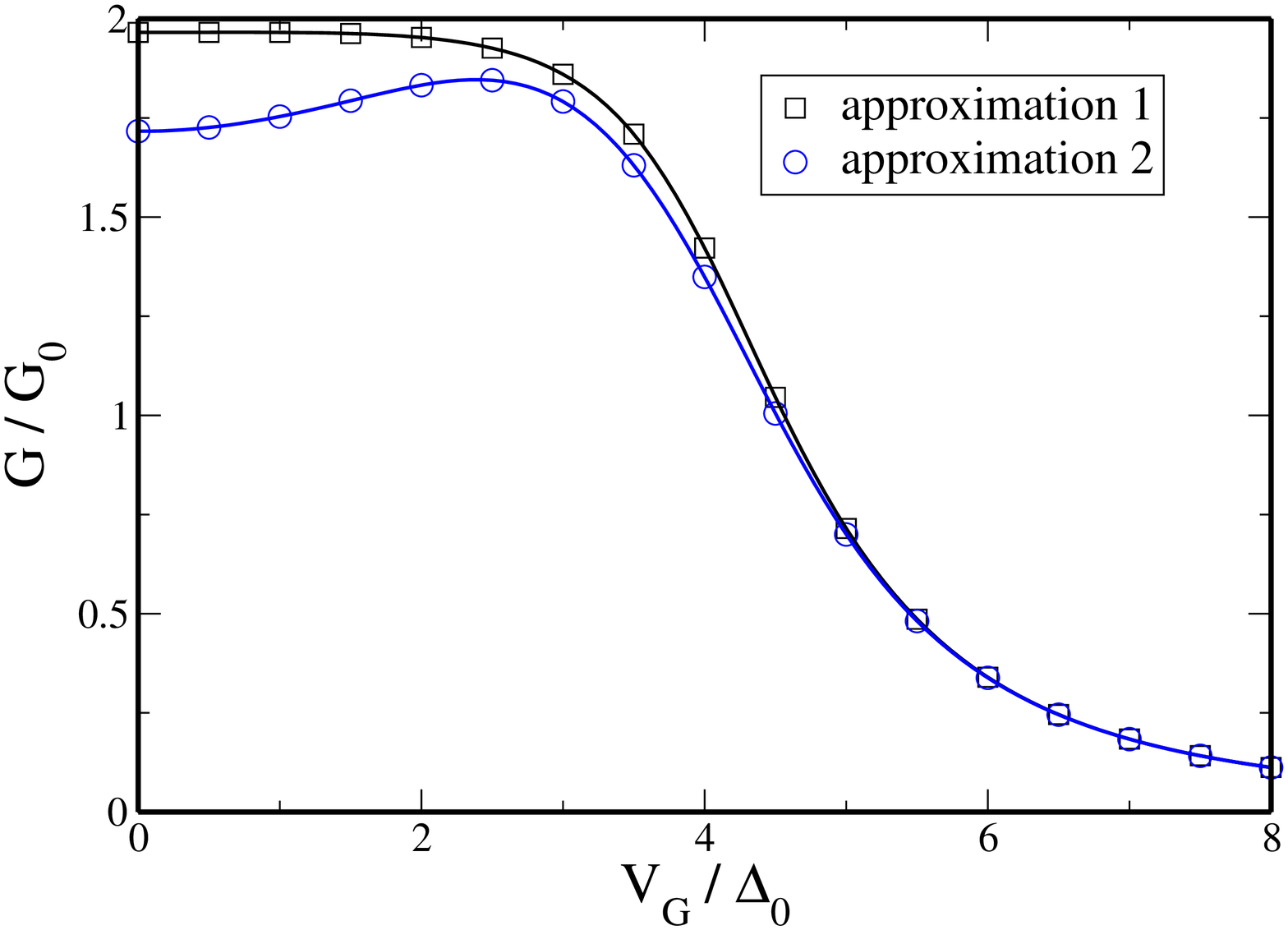}
      \caption[compleitfaehigkeitapprox]{(color online) Comparison of approximation 1 and 2 to the linear conductance for $U=8\Delta_0$, $\beta=50/\Delta_0$.}
      \label{pictcompleitfaehigkeitapprox}
    \end{minipage}
\end{figure}

Furthermore we calculated the linear conductance $G$ from
Eq. (\ref{eqleitfaehigkeit2}). In Fig. \ref{pictcompleitfaehigkeittemp} we
show $G$ as function of the gate-voltage $V_g$ for several temperatures and $U = 8 \Delta_0$. At
low temperatures $\beta=50/\Delta_0$ we get a plateau in the conductance for 
gatevoltages between $-\frac{U}{2}$ and $\frac{U}{2}$, which is due to the
pinning of spectral weight at the Fermi energy. The plateau-value is given by
the unitary limit $2 G_0 = 2 e^2 / h$. For higher temperatures the conductance
at $V_g=0$  decreases quadratically with the temperature.

In Fig. ~\ref{pictcompleitfaehigkeitapprox} the linear conductance derived in
the two approximation schemes 1 and 2 is shown. In approximation 2 the linear
conductance for small gate-voltages is reduced in  
comparison with approximation 1. We understand this again as a
finite-temperature effect. In approximation 2, the Kondo peak gets narrower,
i.e. the effective Kondo scale comes out smaller. Hence, in this approximation
the actual temperature $\beta^{-1}$ is closer to $T_K$ as in approximation 1
and the conductivity shows a stronger finite-temperature suppression. 

\begin{figure}[htbp]
    \centering
    \begin{minipage}[t]{8 cm}
      \includegraphics[width=1.0\textwidth]{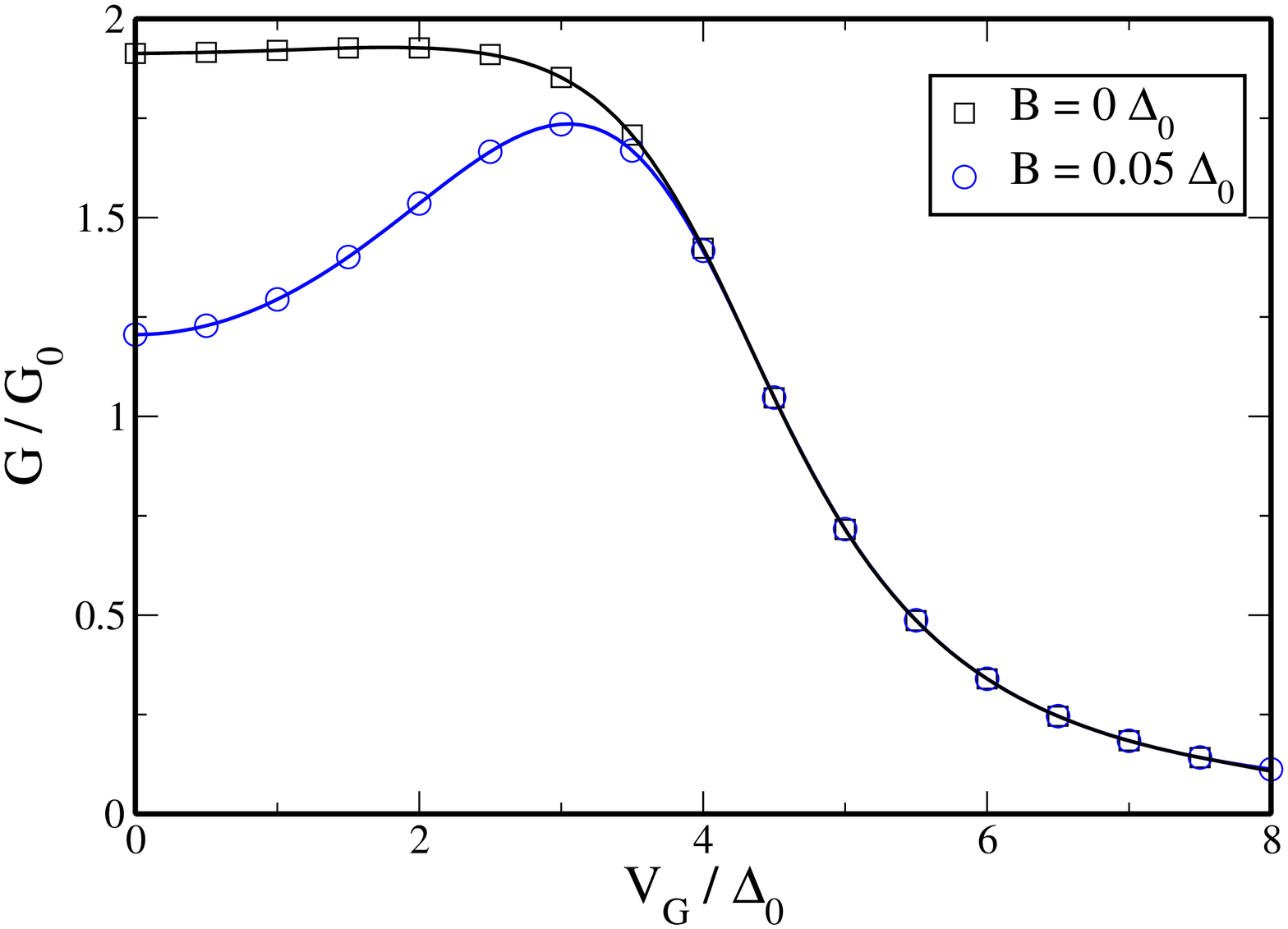}
      \caption[compleitfaehigkeitmagnetfeld]{(color online) Linear conductance for $U=8\Delta_0$, $\beta=30/\Delta_0$ and several values of the magneticfield, calculated in approximation 1.}
      \label{pictcompleitfaehigkeitmagnetfeld}
    \end{minipage}
    \begin{minipage}[t]{8 cm}
      \includegraphics[width=1.0\textwidth]{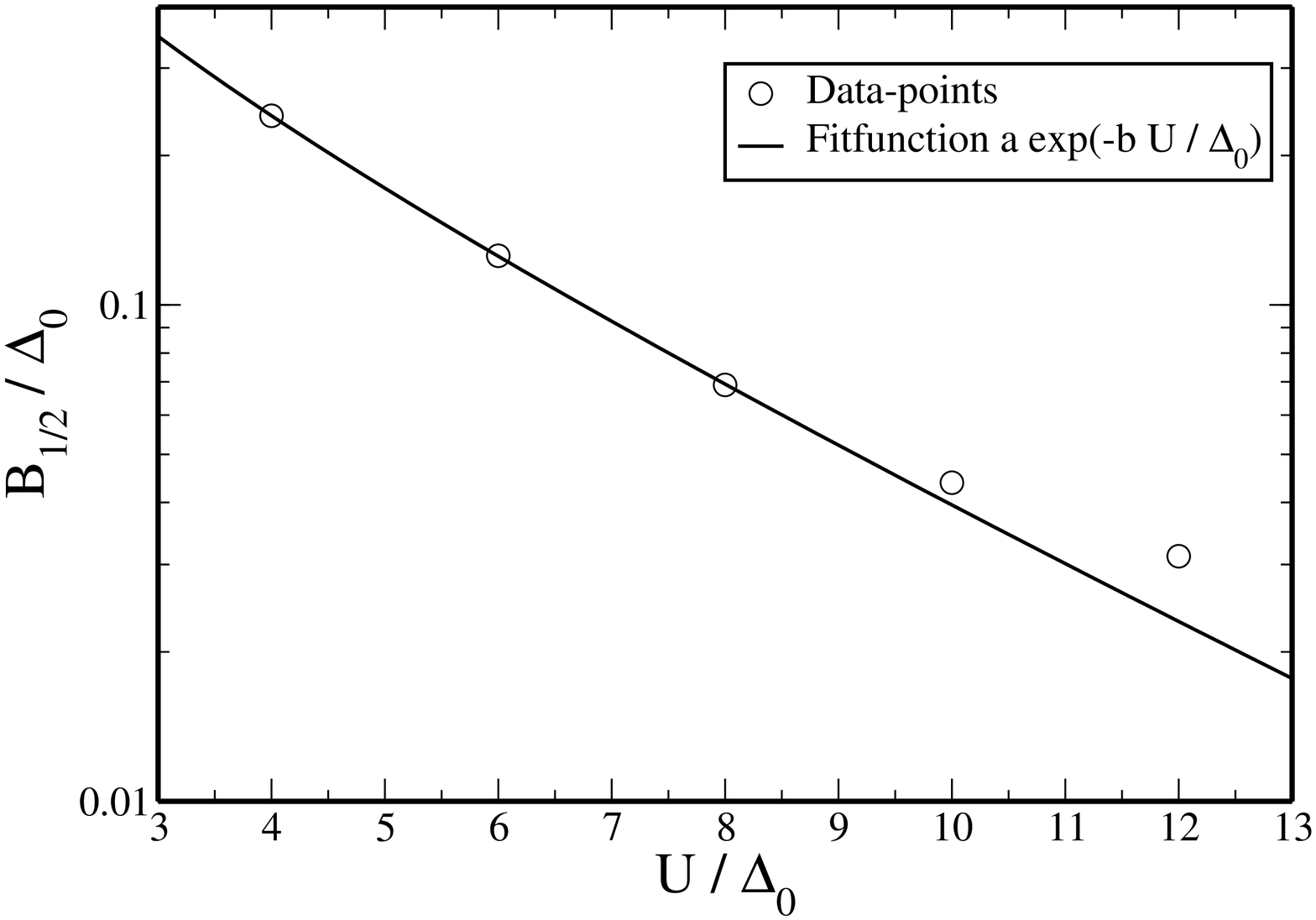}
      \caption[susceptibility]{(color online) $B_{1/2}$ as function of $U$, $\beta=30/\Delta_0$, approximation 1, together with an exponential fit curve.}    
      \label{pictbhalfvsu}
    \end{minipage}
\end{figure}

Fig.~ (\ref{pictcompleitfaehigkeitmagnetfeld}) shows the suppression of the
gate-voltage at $V_G = 0$ due to a finite magnetic field. As shown in
Ref. \onlinecite{Kar06} one can extract the Kondo-scale from 
this supression within a frequency-independent fRG-scheme with
frequency-cutoff. Therefore one defines the Kondo scale $T_K$  as equal to the
magnetic field $B_{1/2}$ that is required to suppress the gate voltage 
$G(V_G=0)$ to $G_0=e^2/h$, which is one half of the unitary limit. In
Fig. (\ref{pictbhalfvsu})  we show $B_{1/2}$ as function of $U$. As shown the
data for small $U$ can be fitted to an exponential curve 
of the form $a \exp\left(-b\, U/\Delta_0\right)$. This behaviour is expected
in the Kondo-regime. Here we find it already for these intermediate values of
$U$. For larger $U$ there are systematic deviations from exponential
behaviour. These deviations begin at $U \sim 8-9 \Delta_0$, where the Kondo
scale according to this association becomes comparable to the temperature,
$T_K \approx \frac{1}{\beta}=\frac{\Delta_0}{30}$. From our fit we get $b
\approx 0.32$, in good agreement with the exact value $b=\pi/8\approx0.39$. \cite{Hew93} 

\subsection{Results for the magnetic suseptibility and Wilson ratio}
We also calculated the static magnetic susceptibility which is defined by
\begin{equation}
\chi_s = \frac{d\left(\langle n_{\uparrow}\rangle -
    \langle n_{\downarrow}\rangle\right)}{d B}\bigg|_{B=0}.  
\end{equation}
Here $\langle n_{\sigma}\rangle$ is the average occupation of electrons with
spin $\sigma$, which is calculated by 
\begin{equation}
\langle n_{\sigma}\rangle=\frac{1}{2}+2 T
\sum_{\alpha>0}R_{\alpha}\text{Re}\mathcal{G}_{\sigma}(i\tilde{\omega}_{\alpha}) 
\end{equation}
with the same $R_{\alpha}$ and $\tilde{\omega}_{\alpha}$ as in Eq. \ref{eqleitfaehigkeit2}.
\begin{figure}[htbp]
    \centering
      \includegraphics[width=0.5\textwidth]{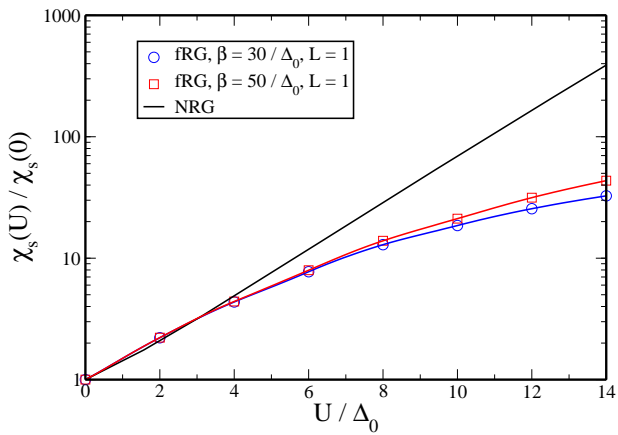}
      \caption[susceptibility]{(color online) Magnetic susceptibility as
        function of $U$, $\beta=30/\Delta_0, 50/\Delta_0$, approximation 1 in comparison with NRG-data.}     
      \label{pictsusceptibility}
\end{figure}
In Fig.~\ref{pictsusceptibility} we show the spin-susceptibility in comparison
with NRG-data. For large values of $U$ the spin-susceptibility is expected to
be inversely proportional to the Kondo temperature $\chi_s \sim
1/T_{K}$. Therefore one expects an exponential dependence on the interaction 
strength. While the susceptibility definitely rises with increasing $U$, the
exponential behaviour is not found in our fRG-approach. A part of this
deviation might again be a thermal effect, as for $U\gtrsim 8-9\Delta_0$ 
the Kondo temperature falls below $\beta^{-1}$ where the
calculation takes place.  

\begin{figure}[htbp]
    \centering
      \includegraphics[width=0.5\textwidth]{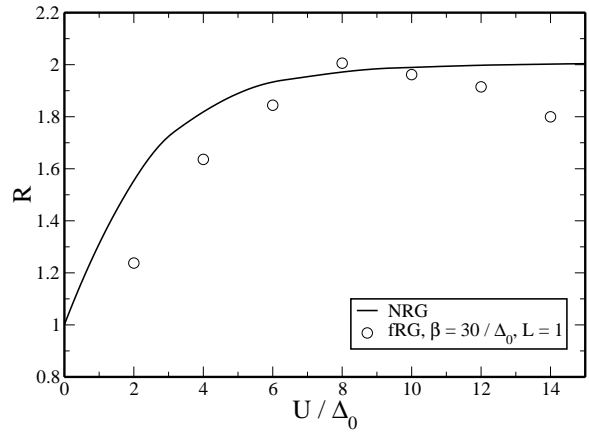}
      \caption[wilsonratio]{Wilson ratio R as function of $U$,
        $\beta=30/\Delta_0$, approximation 1 in comparison with NRG-data.}     
      \label{pictwilsonratio}
\end{figure}

In Fig.~\ref{pictwilsonratio} we show the Wilson ratio which is calculated as
$R=\frac{2 \chi_s}{\chi_s+\chi_c}$ with the charge-susceptibility
$\chi_c=\lim_{\mu\rightarrow 0}\sum_{\sigma} \frac{d\langle
  n_{\sigma}\rangle}{d \mu}$. 
With the relation $\frac{1}{m^{*}}=\frac{4}{\pi\Delta_{0}(\chi_s+\chi_c)}$,
which holds in the Anderson impurity model \cite{Hew93,Kop10} we get
$R=\frac{\Delta_0\pi\chi_s}{2 m^{*}}$. Therefore we can calculate $R$ from our
data of the effective mass and the spin-susceptibility. $R=2$ corresponds to
the Kondo regime, where charge fluctations are completely suppressed and the
charge-susceptibility vanishes. 
As seen in Fig.~\ref{pictwilsonratio}, for  $U\gtrsim 8 \Delta_0$ the fRG-data
come out very close to $R=2$ even though $\chi_s$ is too small in the
fRG. This points to advantageous cancellations of errors for this ratio in  
the fRG. Indeed $m^{*}$ comes out too small as well. The slight decrease of
$R$ for $U > 9 \Delta_0$ might be again due effects of finite temperature. 

\section{Conclusion}
\label{sec:conclusions}
We have presented an alternative renormalization group approach to the
single-impurity Anderson model. The starting point is the exact result for
self-energy and four-point vertex of a small subsystem ('core')
containing the correlated impurity site. Then we track the
evolution of these two functions when the coupling to the bath is switched on
slowly.  This way, the solution
of the small isolated cluster is implemented exactly, and the flow generates
changes of infinite order in the hybridization with the bath. The main
approximation is the truncation of the flow equations after the four-point
vertex. In the present case this means 
that the change of the higher-order vertices (six-point, eight-point, etc.)
upon coupling to the bath is not allowed to influence the lower order
vertices, i.e. the two-particle interaction and the  self-energy. Yet, the
idea that led us take this avenue was that starting with the exact
two-point and four-point vertex of the small core contains enough strong
correlation physics in order to give a physically reasonable results. The
local interacting physics, such as the atomic scales are well represented in
our approach. However, the Kondo effect requires to describe the subtle
interplay with a continuum of states including many energy scales and our
approach captures this only qualitatively, but not quantitatively. 
We expect this to be less of a problem in a self-consistent approach, such as
DMFT, where for stronger coupling the Kondo resonance does not survive.

Indeed, the numerical results for clusters with an odd number of auxiliary
sites ($L=1$ and $L=3$) show that the flow equations produce qualitatively
correct results, whereas for even numbers the Fermi liquid behavior is not
recovered. The dependence of the width of the Kondo peak  with increasing $U$ is
quantitatively different from NRG results, i.e. the correct exponential dependence is
not reproduced. For larger $U$ the Kondo scale gets smaller than the nonzero temperatures for which the fRG scheme is feasible. 
So no clear statement can be made regarding the large-$U$ behavior.
However, in the interesting intermediate coupling regime, the deviations may
be tolerable. In this sense embedding this new impurity solver in a different
contexts to describe itinerant and strongly coupled physics qualitatively
correctly (see discussion below) seems a viable possibility.  

In comparison with other finite-frequency functional RG techniques, e.g. those that are
perturbative in $U$, our data end up in the same range, as shown in
Fig.~\ref{picteffectivemass2}. As the ground state for weak and strong
coupling remains the same, it is not entirely surprising that approaches
starting at the opposite ends lead to qualitatively similar answers. The
quantitative agreement could however be interpreted further, as a measure of
what is missed by the truncation after the four-point vertex that is common to
both lines of approach. Note, that in a recent paper Streib et al. \cite{Stre12} were able to
reproduce the exponential Kondo-scale in a fRG-scheme with partial bosonization of the
transverse spin-fluctations. By using Ward identities they are able to avoid 
further truncations of the flow equations. In this way they obtain the spin-susceptibility and the
effective mass in good agreement with the exact Bethe Ansatz solution.

While the use of the fRG method as impurity solver is for most aspects not
superior to the established techniques, we hope that 
generalizations for correlated lattice systems (i.e. with more than one
correlated site, like the two-dimensional Hubbard model) will be feasible. In
this case, both self-energy and interaction vertex will become increasingly non-local
during the flow, and certainly, suitable approximations have to be found in
order to keep the amount of information manageable. For example, small
correlated cluster cores can be coupled together during the flow via switching
on the hopping amplitude between the clusters from 0 to the original value. The
solution of the core will then provide the spectral weight transfers on the
energy scale $U$ and the accompanying reduction of the spectral weight near
the Fermi level. Together with the core interaction vertex this spectrum will
serve as effective action of a strongly correlated Fermi liquid, which then
can undergo long-range ordering transition when the cores are coupled
together.  Note that in extension of earlier ideas in the vein of cluster perturbation theory (see, e.g. Refs. \onlinecite{Gro93,Sen02}) the fRG scheme also 
allows one to determine the non-local hybridization effects on the interaction, which has direct consequences 
on the character and scale of low-temperature instabilities such as unconventional superconductivity. 
This way we hope to extend the successful functional RG instability
analysis for weakly correlated fermions to the more strongly correlated
regime. The high-energy physics of a strongly interacting Hubbard-like
system is certainly more local than the low-energy physics of collective
ordering. Hence, the 
break-up into small cores and subsequent coupling together also closely
follows the physical intuition of first solving the problem with the largest
energy scale before the low-energy end is considered. 

We acknowledge useful discussions with David Joerg, Manfred Salmhofer, Sabine
Andergassen, Severin Jakobs, Christoph Karrasch, Volker Meden, Andrej Katanin,
Walter Metzner, David Rosen and Walter Hofstetter. 
This project was supported by the DFG research units FOR 732 and FOR 912.
JB acknowledges financial support from the DFG through BA 4371/1-1.
\newpage
\appendix
\section{Expressions for the Green's functions}
\label{sec:appendixgreenfunction}

The inverse free Green's function
$Q\left(i\omega_{n}\right)\equiv\left[\mathcal{G}^{0}\left(i\omega_{n}\right)\right]^{-1}$
on the imaginary frequency axis is given by $Q\left(i\omega_{n}\right)=i\omega_{n}\textbf{1}-\hat{H}_{0}$, where $\hat{H}_{0}$ is the noninteracting part of the
Hamiltonian (\ref{eqhamiltonsiam}).
Written as a matrix $Q\left(i\omega_{n}\right)$ it is given by
\begin{eqnarray}
Q\left(i\omega_{n}\right)&=&\left[\begin{array}{c|c}  Q_{\uparrow}\left(i\omega_{n}\right) & 0   
\\ \hline 0 & Q_{\downarrow}\left(i\omega_{n}\right)
\end{array}\right],
\end{eqnarray}
where
\begin{eqnarray}
Q_{\sigma}\left(i\omega_{n}\right)&=&\left[\begin{array}{c|cccc} & d & b_{1} & b_{2} & \cdots  
\\ \hline d & i\omega_{n}-\epsilon_{d,\sigma} & v & & 
\\  b_{1} & v & i\omega_{n} & t &  
\\  b_{2} & & t & i\omega_{n} & \cdots  
\\  \cdots & & & \cdots & \cdots 
\end{array}\right] \, . 
\end{eqnarray}

The inverse of $Q_{\sigma}\left(i\omega_{n}\right)$ can be calculated by using the
identity
\begin{eqnarray}\label{eqmatrixidentity}
&&\left[\begin{array}{c|c} A & B  
\\ \hline C & D 
\end{array}\right]^{-1}
= \\
\nonumber
&&
\left[\begin{array}{c|c} \left(A-BD^{-1}C\right)^{-1} & -(A-BD^{-1}C)^{-1}BD^{-1}  
\\ \hline -D^{-1}C\left(A-BD^{-1}C\right)^{-1} & (D-CA^{-1}B)^{-1} 
\end{array}\right],
\end{eqnarray}
which is valid for arbitrary invertible matrices $A$, $B$, $C$ and $D$.

If we introduce the matrix $g_{b}\left(i\omega_{n}\right)$ by
\begin{eqnarray}
g_{b}^{-1}\left(i\omega_{n}\right)=\left[\begin{array}{c|cccc}  & b_{1} & b_{2} & b_{3} & \cdots  
\\  \hline b_{1} & i\omega_{n} & t & & 
\\  b_{2} & t & i\omega_{n} & t & 
\\  b_{3} & & t & i\omega_{n} & \cdots  
\\  \cdots & & & \cdots & \cdots 
\end{array}\right]  \, , \label{eqgbinv}
\end{eqnarray}
the free Green's function on the dot site follows from the identity (\ref{eqmatrixidentity}) as
\begin{equation}
\mathcal{G}^{0}_{\sigma}\left(i\omega_{n},d,d\right)=\left(i\omega_{n}-\epsilon_{d,\sigma}-v^2 g_{b}\left(i\omega_{n},b_{1},b_{1}\right)\right)^{-1} \, . 
\end{equation}
The function $g_{b}\left(i\omega_{n},b_{1},b_{1}\right)$ can be calculated again from the identity (\ref{eqmatrixidentity}), this time applied to Eq. (\ref{eqgbinv}), as
\begin{equation}\label{eqfunctionfirstbathsite}
g_{b}\left(i\omega_{n},b_{1},b_{1}\right)=\left(i\omega_{n}-t^2 g_{b}\left(i\omega_{n},b_{1},b_{1}\right)\right)^{-1} \, . 
\end{equation}
Here we used that adding or removing the first bath-site from a semi-infinite
tight-binding-chain do not change the chain. 
Eq. (\ref{eqfunctionfirstbathsite}) can be solved to give the explicit
expression for $g_{b}\left(i\omega_{n},b_{1},b_{1}\right)$ in Eq.(\ref{eqbadmatsubaragreenfunktion}).

By the Dyson equation the full Green's function
is related to $Q_{\sigma}\left(i\omega_{n}\right)$ and the self-energy
\begin{widetext}
\begin{eqnarray}
\left[\mathcal{G}_{\sigma}\left(i\omega_{n}\right)\right]^{-1}&=&Q_{\sigma}\left(i\omega_{n}\right)-\Sigma_{\sigma}\left(i\omega_{n}\right)\\
\nonumber
&=&\left[\begin{array}{c|cccc} & d & b_{1} & b_{2} & \cdots  
\\ \hline d & i\omega_{n}-\epsilon_{d,\sigma}-\Sigma_{d,\sigma}\left(i\omega_{n}\right) & v & & 
\\  b_{1} & v & i\omega_{n} & t & 
\\  b_{2} & & t & i\omega_{n} & \cdots 
\\  \cdots & & & \cdots & \cdots 
\end{array}\right] \, . 
\end{eqnarray} 

Inverting this matrix with the identity (\ref{eqmatrixidentity}) gives the full Green's function on the dot $\mathcal{G}_{\sigma}\left(i\omega_{n},d,d\right)$ in 
Eq.(\ref{eqfullgreensfunktiondotsite}).

In the same way one gets the Green's function for the first and second bath-site, 
\begin{eqnarray}
&&\mathcal{G}_{\sigma}\left(i\omega_{n},b_1,b_1\right)
=\left(i\omega_{n}-\frac{v^2}{i\omega_{n}-\epsilon_{d,\sigma}-\Sigma_{d,\sigma}\left(i\omega_{n}\right)}-t^2
  g_{b}\left(i\omega_{n},b_{1},b_{1}\right)\right)^{-1} \,
, \label{eqgreensfunctionfirstbathsite}\\[1mm] 
&&\mathcal{G}_{\sigma}\left(i\omega_{n},b_2,b_2\right)
=\left(i\omega_{n}-\frac{t^2}{i\omega_{n}-\frac{v^2}{i\omega_{n}-\epsilon_{d,\sigma}-\Sigma_{d,\sigma}\left(i\omega_{n}\right)}}-t^2
  g_{b}\left(i\omega_{n},b_1,b_1\right)\right)^{-1} \,
. \label{eqgreensfunctionsndbathsite} 
\end{eqnarray}
For the other bath-sites with site index $> 2$, analogous expressions can be derived.

\section{Relation between bath and dot self-energy}
\label{sec:appendixrelation}
 
From the identities of the Green's functions in section \ref{sec:effbath2} we can derive the following
relations for the self-energy for $L=0,1,2,3$,
\begin{eqnarray}
L=0:&\
\Sigma_{d,\sigma}\left(i\omega_n\right)&=i\omega_{n}-\epsilon_{d,\sigma}-\frac{(\Lambda v)^2}{\Sigma_{b,\sigma}\left(i\omega_{n}\right)+(\Lambda v)^2\
  \mathcal{G}^{c,(1)}_{core,\sigma}\left(i\omega_{n},b_0,b_0\right)}\,
, \label{eqrelationdotbathl0}\\ 
L=1:&\ \Sigma_{d,\sigma}\left(i\omega_n\right)&=i\omega_{n}-\epsilon_{d,\sigma}-\frac{v^2}{i\omega_{n}-\frac{(\Lambda t)^2}{\Sigma_{b,\sigma}\left(i\omega_{n}\right)+(\Lambda t)^2\ \mathcal{G}^{c,(1)}_{core,\sigma}\left(i\omega_{n},b_1,b_1\right)}} \, , \label{eqrelationdotbathl1}\\
L=2:&\ \Sigma_{d,\sigma}\left(i\omega_n\right)&=i\omega_{n}-\epsilon_{d,\sigma}-\frac{v^2}{i\omega_{n}-\frac{t^2}{i\omega_{n}-\frac{(\Lambda t)^2}{\Sigma_{b,\sigma}\left(i\omega_{n}\right)+(\Lambda t)^2\ \mathcal{G}^{c,(1)}_{core,\sigma}\left(i\omega_{n},b_2,b_2\right)}}} \, , \label{eqrelationdotbathl2}\\
L=3:&\ \Sigma_{d,\sigma}\left(i\omega_n\right)&=i\omega_{n}-\epsilon_{d,\sigma}-\frac{v^2}{i\omega_{n}-\frac{t^2}{i\omega_{n}-\frac{t^2}{i\omega_{n}-\frac{(\Lambda t)^2}{\Sigma_{b,\sigma}(i\omega_{n})+(\Lambda t)^2\mathcal{G}^{c,(1)}_{core,\sigma}(i\omega_{n},b_3,b_3)}}}} \, . \label{eqrelationdotbathl3}
\end{eqnarray}
\end{widetext}

\section{Solution of the 'core'-problem}
\label{sec:appendixcore}

To solve the local core-problem, we diagonalize the 'core'-Hamiltonian exactly. From this solution we derive the local correlation functions using a Lehmann-representation.
For the one-particle-Green's function this representation is given by
\begin{eqnarray}
\nonumber
\mathcal{G}^{(1)}\left(i\omega_{n},i,j\right)&=&\frac{1}{\mathcal{Z}}\sum_{m,n}\frac{\exp\left[-\beta E_{m}\right]+\exp\left[-\beta E_{n}\right]}{i\omega_{n}-(E_n-E_m)}\\
&&\times\langle n|c_{i}|m\rangle \langle n|c_{j}|m\rangle^{*}
\end{eqnarray}
The Lehmann-representation of the two-particle-Green's function is derived in
Ref. \onlinecite{Haf09}. It is given by 
\begin{widetext}
\begin{eqnarray}\label{eqlehmannrepresentationvierpunktfunktion}
\mathcal{G}^{(2)}\left(i_1',\sigma,i\omega_1';i_2',\sigma',i\omega_2'|i_1,\sigma,i\omega_1;i_2,\sigma',i\omega_2\right)&=&\frac{1}{\mathcal{Z}}\sum_{i,j,k,l}\sum_{\Pi}
\phi\left(E_i,E_j,E_k,E_l,i\omega_{\Pi_1},i\omega_{\Pi_2},i\omega_{\Pi_3}\right)\nonumber\\
&&\times\text{sgn}\left(\Pi\right)\langle i|\mathcal{O}_{\Pi_1}|j\rangle\langle j|\mathcal{O}_{\Pi_2}|k\rangle\langle k|\mathcal{O}_{\Pi_3}|l\rangle
\langle l|c_{i_2,\sigma'}|i\rangle\nonumber\\
&&\times\delta_{\omega_1'+\omega_2'+\omega_1+\omega_2,0} .
\end{eqnarray}
Here the frequencies corresponding to creation and annihilation operators have
the same sign. The operators $\mathcal{O}_{i}$  
are defined by $\mathcal{O}_{1}=c^{\dag}_{i_1',\sigma}$, $\mathcal{O}_{2}=c^{\dag}_{i_2',\sigma'}$ and $\mathcal{O}_{3}=c_{i_1,\sigma}$.
The function $\phi$ is given by
\begin{eqnarray}
\phi\left(E_i,E_j,E_k,E_l,i\omega_1,i\omega_2,i\omega_3\right)&=&\frac{1}{i\omega_3+E_k-E_l}\bigg[\frac{1-\delta_{\omega_2,-\omega_3}\delta_{E_j,E_l}}{i\left(\omega_2+\omega_3\right)+E_j-E_l}\nonumber\\
&&\times \left(\frac{e^{-\beta E_i}+e^{-\beta E_j}}{i\omega_1+E_i-E_j}-\frac{e^{-\beta E_i}+e^{-\beta E_l}}{i\left(\omega_1+\omega_2+\omega_3\right)+E_i-E_l}\right)+\delta_{\omega_2,-\omega_3}\delta_{E_j,E_l}\nonumber\\
&&\times \left(\frac{e^{-\beta E_i}+e^{-\beta E_j}}{\left(i\omega_1+E_i-E_j\right)^2}-\beta\frac{e^{-\beta E_j}}{i\omega_1+E_i-E_j}\right)-\frac{1}{i\omega_2+E_j-E_k}
\bigg(\frac{e^{-\beta E_i}+e^{-\beta E_j}}{i\omega_1+E_i-E_j}\nonumber\\
&&-\left(1-\delta_{\omega_1,-\omega_2}\delta_{E_i,E_k}\right)\frac{e^{-\beta E_i}-e^{-\beta E_k}}{i\left(\omega_1+\omega_2\right)+E_i-E_k}+\beta e^{-\beta E_i}\delta_{\omega_1,-\omega_2}\delta_{E_i,E_k}\bigg)\bigg]
\end{eqnarray}
From (\ref{eqlehmannrepresentationvierpunktfunktion}) one gets the connected two-particle-Green's function from the relation
\begin{eqnarray}
\nonumber
\mathcal{G}^{c,(2)}\left(1',2'|1,2\right)&=&\mathcal{G}^{(2)}\left(1',2'|1,2\right)
\nonumber
-\beta\ \mathcal{G}^{(1)}\left(1'|1\right)\mathcal{G}^{(1)}\left(2'|2\right)
+\beta\ \mathcal{G}^{(1)}\left(1'|2\right)\mathcal{G}^{(1)}\left(2'|1\right).
\end{eqnarray}

\section{fRG-flow-equations}
\label{sec:appendixfrg}

Because the action (\ref{eqeffectiveactionumparametrisiert}) is spin-rotation-invariant, the fRG-flow-equations (\ref{eqflussgleichungselbstenergie}) and (\ref{eqflussgleichungvertex})
can be further simplified. Using the spin-conservation, the two-particle vertex is given by
\begin{eqnarray}
\nonumber
\Gamma_{b}^{\Lambda}(i\omega_1',\sigma_1';i\omega_2',\sigma_2'|i\omega_1,\sigma_1;i\omega_2,\sigma_2)
\nonumber
= V_{b}^{\Lambda}(i\omega_1',i\omega_2'|i\omega_1,i\omega_2)\delta_{\sigma_1,\sigma_1'}\delta_{\sigma_2,\sigma_2'}
-\bar{V}_{b}^{\Lambda}(i\omega_1',i\omega_2'|i\omega_1,i\omega_2)\delta_{\sigma_1,\sigma_2'}\delta_{\sigma_2,\sigma_1'}.
\label{eqrelationgammakopplung}
\end{eqnarray}
From the antisymmetry of $\Gamma_{b}^{\Lambda}\left(1',2'|1,2\right)$ under the permutations $1'\leftrightarrow 2'$ and $1\leftrightarrow 2$ it follows that the 
functions $V_b^{\Lambda}$ and $\bar{V}_{b}^{\Lambda}$ obey the relation
\begin{eqnarray}
V_{b}^{\Lambda}\left(i\omega_1',i\omega_2'|i\omega_1,i\omega_2\right)&=&\bar{V}_{b}^{\Lambda}\left(i\omega_2',i\omega_1'|i\omega_1,i\omega_2\right)\nonumber 
=\bar{V}_{b}^{\Lambda}\left(i\omega_1',i\omega_2'|i\omega_2,i\omega_1\right).
\end{eqnarray}
Using this parametrization we get the flow-equations
\begin{eqnarray}
\frac{d}{d\Lambda}\Sigma^{\Lambda}_{b}\left(i\omega\right)&=&-\frac{1}{\beta}\sum_{i\omega'}S^{\Lambda}\left(i\omega'\right)
\left(2 V_{b}^{\Lambda}\left(i\omega,i\omega'|i\omega,i\omega'\right)-V_{b}^{\Lambda}\left(i\omega,i\omega'|i\omega',i\omega\right)\right)\label{eqflussgleichungselbstenergie2}\\
\frac{d}{d\Lambda}V^{\Lambda}_{b}\left(i\omega_1',i\omega_2'|i\omega_1,i\omega_2\right)&=&\Phi^{\Lambda}_{pp}\left(i\omega_1',i\omega_2'|i\omega_1,i\omega_2\right)
+\Phi^{\Lambda}_{dph}\left(i\omega_1',i\omega_2'|i\omega_1,i\omega_2\right)+\Phi^{\Lambda}_{crph}\left(i\omega_1',i\omega_2'|i\omega_1,i\omega_2\right)\label{eqflussgleichungzweiteilchenvertex}
\end{eqnarray}
with
\begin{eqnarray}
\Phi^{\Lambda}_{pp}\left(i\omega_1',i\omega_2'|i\omega_1,i\omega_2\right)&=&\frac{1}{\beta}\sum_{i\omega_3,i\omega_4}L\left(i\omega_3,i\omega_4\right)
V_{b}^{\Lambda}\left(i\omega_3,i\omega_4|i\omega_1,i\omega_2\right) V_{b}^{\Lambda}\left(i\omega_1',i\omega_2'|i\omega_3,i\omega_4\right)\\
\Phi^{\Lambda}_{dph}\left(i\omega_1',i\omega_2'|i\omega_1,i\omega_2\right)&=&-\frac{1}{\beta}\sum_{i\omega_3,i\omega_4}L\left(i\omega_3,i\omega_4\right)
\bigg( 2 V_{b}^{\Lambda}\left(i\omega_1',i\omega_3|i\omega_1,i\omega_4\right) V_{b}^{\Lambda}\left(i\omega_2',i\omega_4|i\omega_2,i\omega_3\right)\nonumber\\
&&- V_{b}^{\Lambda}\left(i\omega_1',i\omega_3|i\omega_1,i\omega_4\right) V_{b}^{\Lambda}\left(i\omega_2',i\omega_4|i\omega_2,i\omega_3\right)\nonumber\\
&&- V_{b}^{\Lambda}\left(i\omega_1',i\omega_3|i\omega_4,i\omega_1\right) V_{b}^{\Lambda}\left(i\omega_2',i\omega_4|i\omega_2,i\omega_3\right)\bigg)\\
\Phi^{\Lambda}_{crph}\left(i\omega_1',i\omega_2'|i\omega_1,i\omega_2\right)&=&\frac{1}{\beta}\sum_{i\omega_3,i\omega_4}L\left(i\omega_3,i\omega_4\right)
V_{b}^{\Lambda}\left(i\omega_2',i\omega_3|i\omega_4,i\omega_1\right) V_{b}^{\Lambda}\left(i\omega_1',i\omega_4|i\omega_3,i\omega_2\right).
\end{eqnarray}
The function $L$ is defined as
\begin{equation}
L\left(i\omega_1,i\omega_2\right)=\mathcal{G}^{\Lambda}\left(i\omega_1\right)S^{\Lambda}\left(i\omega_2\right)
+\mathcal{G}^{\Lambda}\left(i\omega_2\right)S^{\Lambda}\left(i\omega_1\right).
\end{equation}
The single-scale-propagator is given by
\begin{equation}
S^{\Lambda}\left(i\omega\right) = \frac{-2\Lambda\left(i\omega-t^2 g_b\left(i\omega,b_1,b_1\right)\right)}
{\left(i\omega-(t\Lambda)^2\mathcal{G}^{c,(1)}_{core}\left(i\omega,b_L,b_L\right)-t^2 g_b\left(i\omega,b_1,b_1\right)-\Lambda^2\Sigma_b^{\Lambda}\left(i\omega\right)\right)^2}.
\end{equation}
The initial conditions for $\Lambda=0$ are
\begin{eqnarray}
\Sigma_b^{\Lambda=0}\left(i\omega\right)&=&0,\\
\label{eqanfangsbedingungvertex}
V_b^{\Lambda=0}\left(i\omega_1',i\omega_2'|i\omega_1,i\omega_2\right)&=&t^4\mathcal{G}_{core}^{c,(2)}\left(i\omega_1',b_{L},\uparrow;i\omega_2',b_{L},\downarrow|i\omega_1,b_{L},\uparrow;i\omega_2,b_{L},\downarrow\right).
\end{eqnarray}
\end{widetext}

\bibliography{siam}

\end{document}